\documentclass[]{aa}
\usepackage{amsmath}
\usepackage{graphics}
\usepackage{natbib}
\usepackage{amsmath}
\usepackage{amssymb}

\begin{document}
\title{Turbulence/wave transmission at an ICME-driven shock observed by Solar Orbiter and Wind}
\author{L.-L.~Zhao\inst{1}
\and G.~P.~Zank\inst{1,2}
\and J.~S.~He\inst{3}
\and D.~Telloni\inst{4}
\and Q.~Hu\inst{1,2}
\and G.~Li\inst{1,2}
\and M.~Nakanotani\inst{1}
\and L.~Adhikari\inst{1}
\and E.~K.~J.~Kilpua\inst{5}
\and T.~S.~Horbury\inst{6}
\and H.~O'Brien\inst{6}  
\and V.~Evans\inst{6} 
\and V.~Angelini\inst{6}
}

\institute{Department of Space Science, The University of Alabama in Huntsville, Huntsville, AL 35899, USA  
\and Center for Space Plasma and Aeronomic Research (CSPAR), The University of Alabama in Huntsville, Huntsville, AL 35805, USA 
\and School of Earth and Space Sciences, Peking University, Beijing 100871, Beijing, People's Republic of China 
\and National Institute for Astrophysics Astrophysical Observatory of Torino Via Osservatorio 20, 10025 Pino Torinese, Italy  
\and Department of Physics, University of Helsinki, Helsinki, Finland  
\and Imperial College London, South Kensington Campus, SW7 2AZ, UK  
}

\abstract{} 
{\textit{Solar Orbiter} observed an interplanetary coronal mass ejection (ICME) event at 0.8 AU on 2020 April 19. The ICME was also observed by \textit{Wind} at 1 AU on 2020 April 20. An interplanetary shock wave was driven in front of the ICME. We focus on the transmission of the magnetic fluctuations across the shock and analyze the characteristic wave modes of solar wind turbulence in the vicinity of the shock observed by both spacecraft.} 
{The ICME event is characterized by a magnetic helicity-based technique. The ICME-driven shock normal is determined by magnetic coplanarity method for \textit{Solar Orbiter} and using a mixed plasma/field approach for \textit{Wind}. The power spectra of magnetic field fluctuations are generated by applying both a fast Fourier transform and Morlet wavelet analysis. To understand the nature of waves observed near the shock, we use the normalized magnetic helicity as a diagnostic parameter.
The wavelet reconstructed magnetic field fluctuation hodograms are used to further study the polarization properties of waves.
} 
{We find that the ICME-driven shock observed by \textit{Solar Orbiter} and \textit{Wind} is a fast forward oblique shock with a more perpendicular shock angle at \textit{Wind}'s position. After the shock crossing, the magnetic field fluctuation power increases. Most of the magnetic field fluctuation power resides in the transverse fluctuations. In the vicinity of the shock, both spacecraft observe right-hand polarized waves in the spacecraft frame. The upstream wave signatures fall in a relatively broad and low frequency band, which might be attributed to low frequency MHD waves excited by the streaming particles. For the downstream magnetic wave activity, we find oblique kinetic Alfv\'en waves with frequencies near the proton cyclotron frequency in the spacecraft frame. The frequency of the downstream waves increases by a factor of $\sim$7--10 due to the shock compression and the Doppler effect.} 
{} 

\titlerunning{Waves near a shock}
\authorrunning{L.-L.~Zhao et al.}

\maketitle

\section{Introduction}\label{sec:introduction}
Interplanetary shocks in the heliosphere have important consequences for the generation and evolution of solar wind turbulence. However, the direct impact of a shock on its ambient turbulence is not well studied.
The interaction of a shock with turbulence is interesting from several different perspectives. First, large-scale MHD waves interacting with shocks can be modeled as the transmission and reflection of waves at an ideal discontinuity.
In this regard, turbulence is treated as a combination of linear waves that have wavelength much longer than the thickness of the shock.
This problem has been treated by several previous works, such as \cite{mckenzie1968, mckenzie1969, zank2021}.
These studies discussed the transmission of MHD waves such as Alfv\'en waves and in cases of small background magnetic field, vortices and magnetic islands from upstream to downstream.
In general, the wavelength of the transmitted waves tends to be smaller than the upstream waves. The fluctuation power downstream is typically larger than that upstream, which has been verified by satellite observations \citep[e.g.,][]{zank2006, hu2013power, adhikari2016, zhao2019, borovsky2020, Borovsky2020JGR}.
However, this approach may not be applicable when the turbulent nonlinearity is not negligible, especially if the turbulence is strong. In this case, the back reaction of turbulence on shocks needs to be considered, which results in the Rankine-Hugniot jump conditions being modified by turbulence \citep[e.g.,][]{zank2002}, and the amplitude of waves/fluctuations being greatly enhanced as it transmits to the downstream \citep[e.g.,][]{lu2009}.  

One important consequence of the turbulence-shock interaction is its effect on particle transport and acceleration. Self-generated fluctuations upstream of the shock can be amplified upon crossing the shock and these upstream and downstream waves may effectively scatter particles leading to efficient 
diffusive acceleration \citep[e.g.,][]{mckenzie1982, li2003, Li2005AIPC, rice2003, Vainio2007}.
Besides the increase of fluctuation power with the shock crossing, the change in the turbulence properties such as the compressibility and anisotropy will also affect the transport of particles.
For example, the generation of magnetic islands may result in additional particle acceleration due to magnetic reconnection \citep{zank2014, zank2015, le2015kinetic,le2016combining, zhao2018, zhao2019, zhao2019acr, adhikari2019}.

Another aspect of the turbulence-shock interaction is related to kinetic-scale fluctuations. Kinetic-scale fluctuations are commonly observed in the solar wind and they are thought to play an important role in dissipation processes through wave-particle interactions.
Previous observational analysis has identified different types of kinetic waves in the solar wind, which are thought to be responsible for the observed steepening of the magnetic fluctuation spectrum above the ion cyclotron frequency. Different wave modes can be identified based on their polarization properties. For example, ion cyclotron waves (ICWs) propagate nearly parallel to the magnetic field and possess a left-handed polarization in the solar wind frame \citep{jian2009, jian2010, he2011, he2019direct, Bruno2015, telloni2019}, kinetic Alfv\'en waves (KAWs) propagate in a direction quasi-perpendicular to the magnetic field and are right-handed polarized \citep{bale2005, he2011oblique, podesta2013evidence, woodham2018role, telloni2020wave}, and whistler waves are right circularly polarized and propagate quasi-parallel to the magnetic field \citep{gary2009, podesta2011effect, podesta2011magnetic, salem2012, tenbarge2012, Zhu2019}. Whistler waves can propagate obliquely or very perpendicularly, and may coexist with KAWs. However, according to linear kinetic theory, oblique whistler waves tend to have large magnetic compressibility with strong parallel fluctuations $dB_{\parallel}$. In contrast, KAWs are mostly dominated by perpendicular fluctuations $dB_{\perp}>dB_{\parallel}$ \citep{gary2009, he2011oblique, salem2012}.

Although the interaction of kinetic waves and shocks is not very well studied, it is widely accepted that wave-particle interactions provide the primary dissipation mechanism for collisionless shocks. Observations of kinetic waves near interplanetary shocks have been reported by \cite{wilson2009, wilson2016low}, but these studies do not address how these waves evolve with the shock crossing.
In this paper, we study the details of waves and turbulence in the vicinity of an interplanetary shock observed by \textit{Solar Orbiter} and \textit{Wind} on 2020 April 19--20. The shock is driven by an interplanetary coronal mass ejection (ICME), which is also observed by both spacecraft.
Due to \textit{Solar Orbiter} not making plasma measurements during this period, we focus mainly on the transmission of magnetic fluctuation properties, such as the magnetic fluctuation power and compressibility. We also look into possible kinetic-scale wave activity both upstream and downstream of the shock based on the $\theta_\mathrm{B_{0l} R}$ distribution of the normalized magnetic helicity $\sigma_m$ spectra with $\theta_\mathrm{B_{0l} R}$ being the angle between the local mean magnetic field $B_\mathrm{0l}$ \citep{horbury2008} and the radial direction.

The outline of this paper is as follows. Section 2 presents an overview of the large-scale ICME structure and its driven shock observed both by \textit{Solar Orbiter} and \textit{Wind}. 
Section 3 provides the preliminary shock parameters and the magnetic field fluctuation spectra upstream and downstream of the shock observed by both spacecraft. Spectra of the normalized magnetic helicity $\sigma_m$ are also shown as a diagnostic parameter of wave polarization characteristics.
Section 4 shows the distribution of $\theta_\mathrm{B_{0l} R}$ in $\sigma_m$ spectra and the wavelet reconstructed magnetic field fluctuation hodograms in the $T$--$N$ plane to determine the wave modes.
Section 5 provides a summary and discussions.

\section{Observation of the ICME and its driven shock}\label{sec:CME-shock}

\begin{figure}[htbp]
\centering
\includegraphics[width=0.85\linewidth]{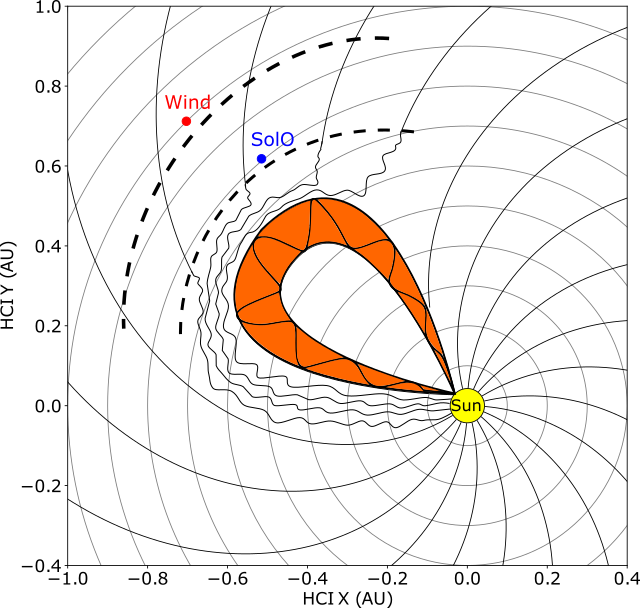} 
	\caption{\textit{Solar Orbiter} (blue dot) and \textit{Wind} (red dot) positions in the $X$-$Y$ plane of the Heliocentric Inertial (HCI) coordinate system at the time when the ICME event of 2020 April was observed by both of them. The schematic is plotted looking down on the ecliptic plane. The Parker magnetic field lines are also shown. The ICME-driven shock (dashed lines) is observed by both spacecraft.}\label{fig:schematic}
\end{figure}
Figure \ref{fig:schematic} illustrates the ICME and its driven shock in the $X$-$Y$ plane of the Heliocentric Inertial (HCI) coordinate system at the time when they were observed.
The locations of \textit{Solar Orbiter} and \textit{Wind} are identified by the blue and red dots.
The orange-colored region represents the ICME flux rope and the dashed lines represent the shock as it approaches the spacecraft.
The Parker spiral magnetic field lines are also shown for reference. 
The shock reached \textit{Solar Orbiter} on 2020 April 19, 05:06:18 UT. \textit{Solar Orbiter} was located then 0.80 AU from the Sun and had an HCI longitude of 130$\degr$ and latitude of -3.94$\degr$. \textit{Wind} observed the shock on 2020 April 20, 01:33:04 UT and was that at $\sim$1.0 AU from the Sun and had an HCI longitude of 134.63$\degr$ and latitude of -5.17$\degr$.
The longitudinal separation between \textit{Solar Orbiter} and \textit{Wind} is around 4.63$\degr$ and the latitude separation is around 1.23$\degr$. Therefore, \textit{Wind} and \textit{Solar Orbiter} are approximately radially aligned during this ICME event, and the radial separation is 0.2 AU.

\begin{figure}[htbp]
\centering
\includegraphics[width=1.0\linewidth]{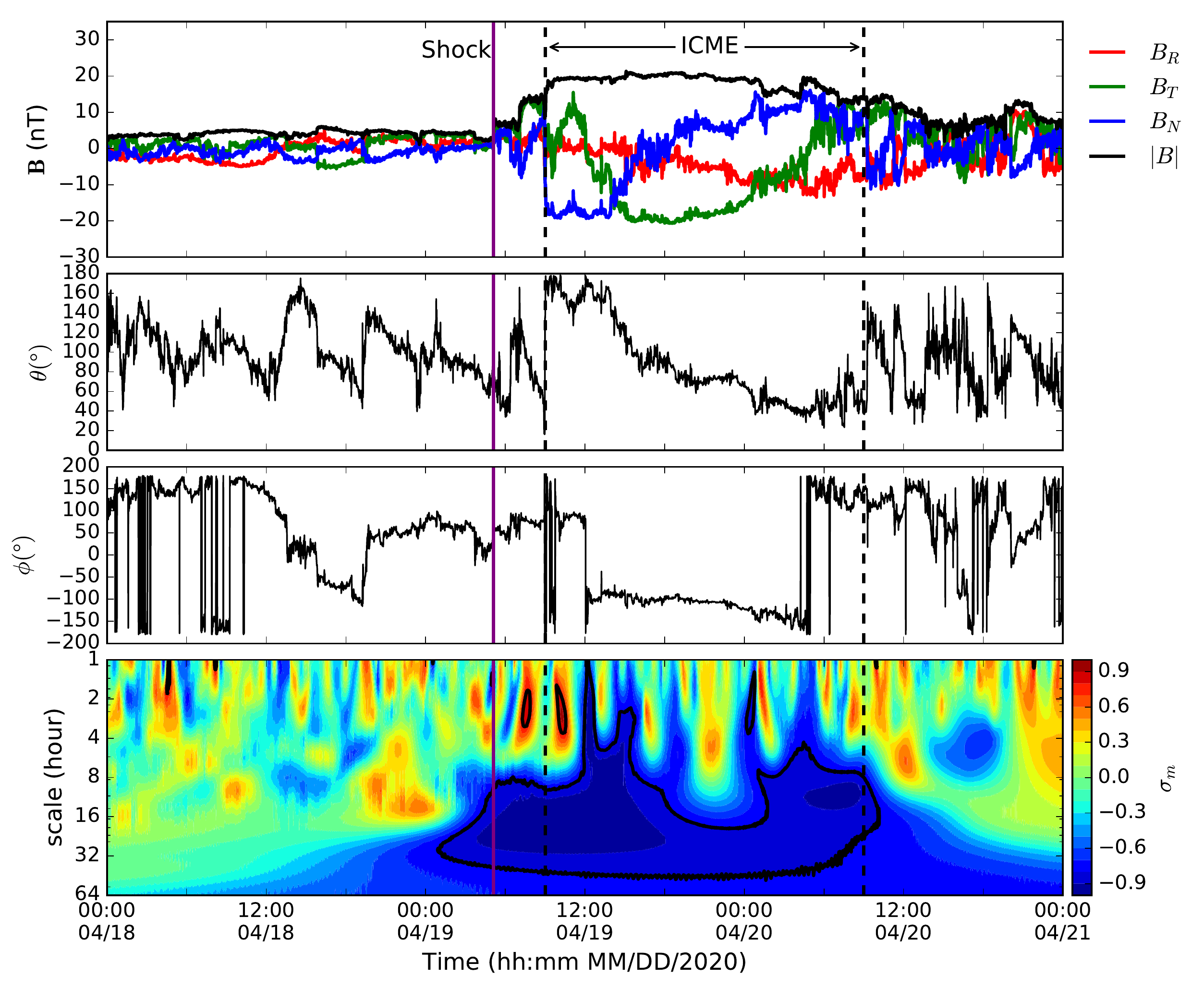} 
	\caption{\textit{Solar Orbiter} observation of the ICME (bounded by the dashed vertical lines) and its driven shock event (solid vertical line) during the period from 2020 April 18 to 2020 April 21. Top panel shows time profiles of the magnetic field vector measured by the \textit{Solar Orbiter}/MAG instrument. The second and third panels show the elevation ($\theta$) and azimuthal ($\phi$) angles of the magnetic field direction in the RTN coordinate system. The bottom panel shows the spectrogram of the normalized magnetic helicity $\sigma_m$ using the Morlet wavelet method. Contour lines are drawn at levels of $|\sigma_m|=0.8$. }\label{fig:solo-CME}
\end{figure}

\begin{figure}[htbp]
\centering
\includegraphics[width=1.0\linewidth]{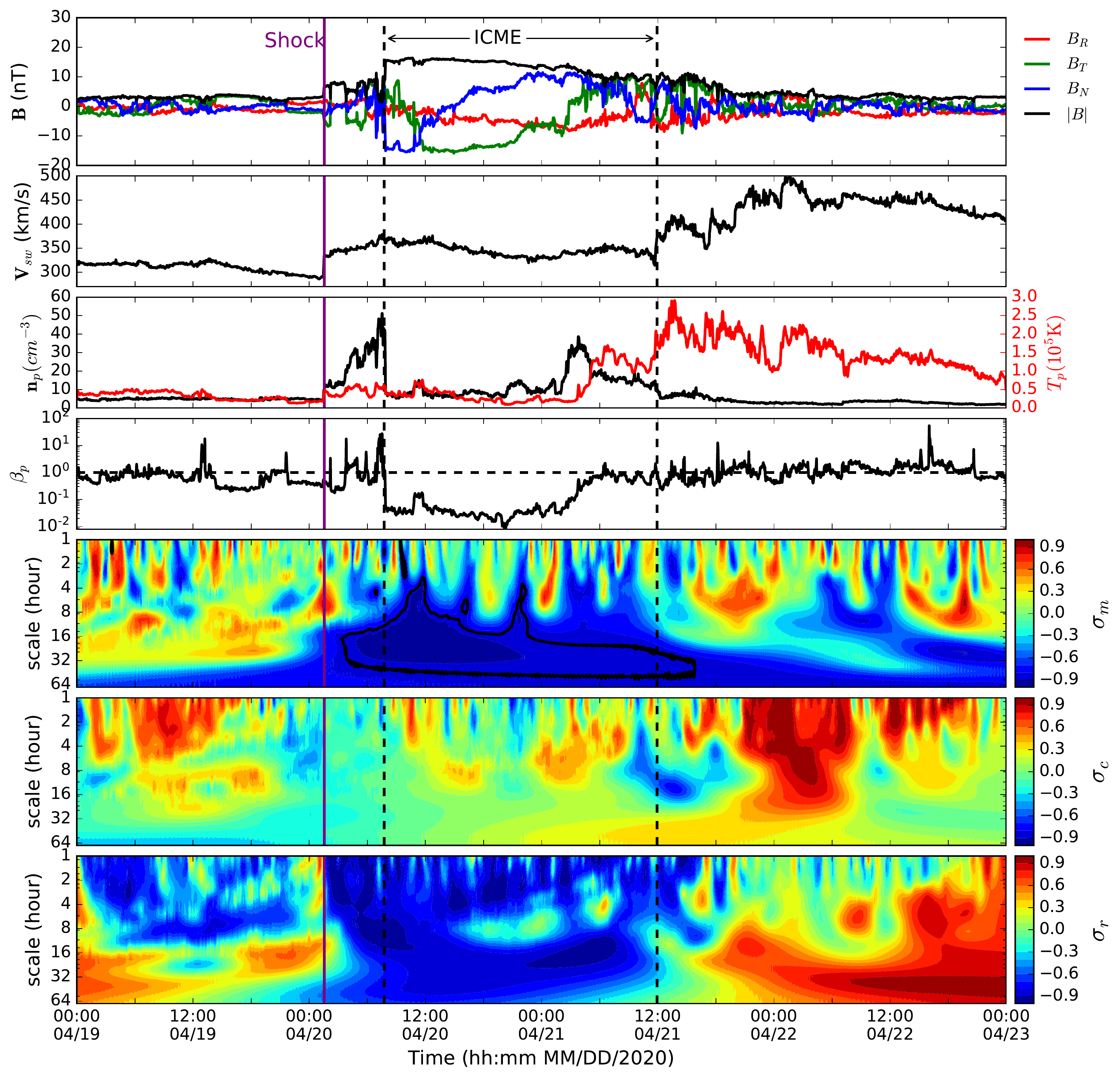} 
\caption{\textit{Wind} observation of the ICME and its driven shock during the period from 2020 April 19 to 2020 April 23. The panels from top to bottom show the magnetic field vector and magnitude, the flow speed, the proton number density and temperature, the proton plasma beta, and spectrograms of the normalized magnetic helicity $\sigma_m$, the normalized cross helicity $\sigma_c$, and the normalized residual energy $\sigma_r$.}\label{fig:wind-CME}
\end{figure}

Figure \ref{fig:solo-CME} is an overview of the ICME event and its driven shock observed by \textit{Solar Orbiter} (SolO) during the period between 2020 April 18 and 2020 April 21. The ICME has been studied in detail in \cite{davies2020} using multi-spacecraft measurements.
Plasma measurements are not available during this period so we show only the magnetic field data measured by the SolO/\textit{MAG} instrument \citep{Horbury2020}.
The top panel shows the magnetic field strength $|B|$ and its three components $B_R$, $B_T$, and $B_N$.
The solid vertical line marks an abrupt increase in $|B|$ that is identified as a forward interplanetary shock crossing.
An ICME is first seen at $\sim$ April 19, 09:00 UT, and lasts about 24 hours. The vertical dashed lines in each panel enclose the observed ICME structure, which is characterized by a smooth magnetic field rotation through a large angle and the enhanced magnetic field strength compared to the surrounding solar wind \citep[e.g.,][]{burlaga1981, kilpua2017}. The averaged magnetic field magnitude within the ICME interval is about 18.4 nT.
The second and third panels show the elevation ($\theta$) and azimuthal ($\phi$) angles of the magnetic field.
The smooth rotation of the magnetic field within the ICME interval can be clearly seen from the elevation angle. 
In the bottom panel, we plot the normalized magnetic helicity $\sigma_{\mathrm{m}}$ \citep{matthaeus1982} calculated by the wavelet method.
In this figure and subsequent analysis, we use the complex Morlet wavelet function
\begin{equation}\label{eq:wavelet}
	\psi(t')= \frac{1}{\sqrt{\pi A}}\exp^{-\frac{t'^2}{A}}\exp^{2\pi i C t'}
\end{equation}
with bandwidth $A = 2.0 $ and center frequency $C =1.0$ Hz; $t'$ is the time normalized by the wavelet scales \citep{Torrence1998}.
The scale and time dependent normalized magnetic helicity $\sigma_{\mathrm{m}}$ can be estimated by 
\begin{equation}\label{sigmam}
	\sigma_m(s,t) = \frac{2 \operatorname{Im}(\tilde{B}_T^*\tilde{B}_N)}{(|\tilde{B}_R|^2 + |\tilde{B}_T|^2 + |\tilde{B}_N|^2)},
\end{equation}
where the tilde represents wavelet transformed quantities, as does the tilde below. $\operatorname{Im}$ denotes the imaginary part of a complex number, $s$ is the wavelet scale and is chosen to be between 1 hour and 64 hours in Fig.\ \ref{fig:solo-CME} and Fig.\ \ref{fig:wind-CME}, and the asterisk represents the complex conjugate. As shown in our previous studies \citep{telloni2020, zhao2020a, zhao2020b}, the ICME, as a large-scale magnetic flux rope, usually possesses a high value of normalized magnetic helicity due to the rotation of the magnetic field over a large angle. The black contour lines
in the panel of the normalized magnetic helicity $\sigma_m$ enclose high magnetic helicity regions with $|\sigma_m| \ge 0.8$.
The ICME observed by \textit{Solar Orbiter} is clearly identified as a left-handed magnetic helical structure ($\sigma_m < 0$). The averaged $\sigma_m$ in the region bounded by the black contour line is around -0.89.  

Figure \ref{fig:wind-CME} shows the \textit{Wind} magnetic field and plasma measurements during the ICME passage. 
The top four panels show the magnetic field magnitude and the three components; the solar wind speed ($V_\mathrm{sw}$); the proton number density ($n_p$) and temperature ($T_p$); and the proton plasma beta ($\beta_p$). 
The forward shock is again indicated by a solid vertical line and is characterized by abrupt increases in the magnetic field strength, solar wind speed, proton density and temperature. The ICME event starts at \textit{Wind} at $\sim$ 07:45 UT and lasts around 28 hours. It shows the typical magnetic cloud signatures, i.e., the abnormally low proton plasma beta due to the enhanced magnetic field strength and smoothly rotating field direction over an interval of a day \citep[e.g.,][]{burlaga1981}, and the low proton temperature. During the ICME interval, the averaged magnetic field magnitude $\langle |B|\rangle \simeq 13.5$ nT, solar wind speed $\langle V_{sw} \rangle \simeq 345$ km/s, proton density $\langle n_{p} \rangle \simeq 12.4$ $cm^{-3}$, proton temperature $\langle T_{p} \rangle \simeq 54434$ K, and the proton plasma beta $\langle \beta_{p} \rangle \simeq 0.24$. The ICME is preceded by a slow solar wind ($\sim$330 km/s). The trailing wind is not particularly fast (peak speed $\sim$500 km/s), but there is a clear positive speed gradient between the ICME and the solar wind behind. The ICME sheath, which is the region between the shock and the ICME ejecta, shows multiple plasma beta jumps that can be related to current sheet crossings \citep[e.g.,][]{li2007, liu2014, huang2016}.
In the bottom three panels, we plot the wavelet spectrograms of the normalized magnetic helicity $\sigma_m$, the normalized cross helicity $\sigma_c$, and the normalized residual energy $\sigma_r$.
The latter two quantities can be calculated from the Els\"assar variables $\boldsymbol{z}^{\pm } = \boldsymbol{u} \pm \boldsymbol{b}/\sqrt{4\pi n_\mathrm{p} m_\mathrm{p}}$ with $\vec{u}$ and $\vec{b}$ the fluctuating velocity and magnetic field vectors, $n_\mathrm{p}$ the proton number density, and $m_\mathrm{p}$ the proton mass \citep{Zank2012}:
\begin{equation}\label{sigmac1}
	\sigma_\mathrm{c}(s,t) = \frac{\langle \tilde{z}^{+2} \rangle - \langle \tilde{z}^{-2}\rangle}{\langle \tilde{z}^{+2} \rangle + \langle \tilde{z}^{-2}\rangle} = \frac{2 \langle \tilde{\vec{u}} \cdot \tilde{\vec{b}}\rangle}{\langle \tilde{u}^2 \rangle + \langle \tilde{b}^2 \rangle},
\end{equation}
and
\[ \sigma_\mathrm{r}(s, t) = \frac{2 \langle \tilde{\boldsymbol{z}}^+ \cdot \tilde{\boldsymbol{z}}^- \rangle}{\langle \tilde{z}^{+2} \rangle + \langle \tilde{z}^{-2} \rangle} = \frac{\langle \tilde{u}^2 \rangle - \langle \tilde{b}^2 \rangle}{\langle \tilde{u}^2 \rangle + \langle \tilde{b}^2 \rangle}
. \]
After the shock passage, the cross helicity $\sigma_\mathrm{c}$ is almost zero, and the residual energy $\sigma_\mathrm{r}$ becomes more negative. 
Within the ICME interval, the averaged $\langle \sigma_\mathrm{m} \rangle \simeq$ -0.9 (left-handed helical structure), $\langle \sigma_\mathrm{c}\rangle \simeq$ 0.07, $\langle \sigma_\mathrm{r} \rangle \simeq$ -0.73. The close-to-zero $\sigma_c$ indicates that there is an almost equal amount of energy propagating parallel and anti-parallel to the magnetic field, i.e., turbulence is balanced. The highly negative $\sigma_\mathrm{r}$ indicates that the energy of the fluctuating magnetic field $\langle b^2 \rangle$ dominates compared to the kinetic fluctuation energy $\langle u^2 \rangle$. These two turbulent properties of the ICME flux rope structures have been widely studied \citep{telloni2020, zhao2020a, zhao2020b, good2020cross}. After the passage of the ICME, \textit{Wind} tends to measure slightly faster solar wind with an increased $\sigma_c$. 

Compared to \textit{Solar Orbiter} observation at 0.8 AU, \textit{Wind} observations at 1 AU suggest that the ICME has expanded slightly as its duration increases from $\sim$24 hours to $\sim$28 hours. The magnetic helicity in the ICME interval is almost unchanged. Due to the lack of plasma data from \textit{Solar Orbiter} in this period, we cannot compare the changes in the cross helicity and residual energy during the evolution of the ICME. The ICME-driven shock observed by both \textit{Solar Orbiter} and \textit{Wind} shows a small jump in the magnetic field magnitude, with the downstream increase being a factor of $\sim$2. However, the ICME sheath observed by both spacecraft shows obvious differences. The sheath observed by \textit{Wind} is more dynamic and has multiple increase in plasma beta.

\section{SolO and Wind observation near the shock}\label{sec:results}

In the following analysis, we focus on the region in the vicinity of the ICME-driven shock observed by \textit{Solar Orbiter} and \textit{Wind}.
The shock parameters calculated at the two locations are summarized in Table 1. 
The shock normal at the \textit{Solar Orbiter} position is obtained by the magnetic coplanarity method \citep{Burlaga1995}:
\begin{equation}\label{eq:mag}
{{\mathbf{\hat n}}_{MC}} =  \pm \frac{{\left( {{{\mathbf{B}}_d} \times {{\mathbf{B}}_u}} \right) \times \Delta {\mathbf{B}}}}{{\left| {\left( {{{\mathbf{B}}_d} \times {{\mathbf{B}}_u}} \right) \times \Delta {\mathbf{B}}} \right|}},
\end{equation}
where $\mathbf{B}_d$ denotes the downstream mean magnetic field, $\mathbf{B}_u$ the upstream mean magnetic field, and $\Delta \mathbf{B} = \mathbf{B}_d-\mathbf{B}_u$.    

The shock normal at the \textit{Wind} position is calculated by a mixed coplanarity method:
\begin{equation}\label{eq:mix}
{{{\mathbf{\hat n}}}_{MX1}} =  \pm \frac{{\left( {{{\mathbf{B}}_u} \times \Delta {\mathbf{V}}} \right) \times \Delta {\mathbf{B}}}}{{\left| {\left( {{{\mathbf{B}}_u} \times \Delta {\mathbf{V}}} \right) \times \Delta {\mathbf{B}}} \right|}},
\end{equation}
where $\Delta \mathbf{V} = \mathbf{V}_d-\mathbf{V}_u$. The speed of the shock observed by \textit{Wind} is estimated by means of the mass flux algorithm using plasma measurements:
\[{V_{sh}} = \frac{{\Delta \left( {\rho {\mathbf{V}}} \right)}}{{\Delta \rho }} \cdot {\mathbf{\hat n}},\]
where $\rho=n_p m_p$ is the proton mass density, $\mathbf{\hat n}$ is the shock normal, and $\Delta \rho = {\rho_d} - {\rho_u}$. Here, quantities with subscripts $u$ and $d$ correspond to their upstream and downstream mean values. The \textit{Solar Orbiter}'s upstream interval for calculating the mean value is from 05:00 to 05:05 UT on April 19, and the downstream interval is between 05:07 and 05:12 UT. The \textit{Wind}'s upstream interval for taking a mean is from 01:15 to 01:30 UT on April 20, and the downstream interval starts from 01:35 to 01:50 UT. These intervals are chosen on the basis that they exclude shock layers and do not include non-shock related disturbances, but are long enough to average out the turbulence and wave activities. In Table 1, 
the rows from top to bottom list the shock normal direction $\bf{\hat n}$, shock obliquity $\Theta_{Bn}$ (the angle between upstream mean magnetic field and the shock normal), shock speed $V_{sh}$, upstream solar wind speed $V_u$, upstream magnetic field $B_u$, velocity jump $V_d/V_u$, magnetic jump $B_d/B_u$, flow speed changes along the shock normal $|\Delta \bf{V} \cdot \bf{\hat n}|$, proton density jump $n_{pd}/n_{pu}$, temperature jump $T_{pd}/T_{pu}$, upstream Alfv\'en speed $V_{Au}$, upstream fast magnetosonic speed $V_{fu}$, upstream proton plasma beta $\beta_{pu}$, and upstream fast mode mach number $M_{fu}$.  
As show in the table, the normals for both shock indicate that the shock front is (at least locally), almost perpendicular to the Sun-Earth line. The shock is quasi-perpendicular at \textit{Wind}'s position with $\Theta_{Bn}=72\degr$ form (\ref{eq:mix}) and 71$\degr$ from (\ref{eq:mag}), while at \textit{Solar Orbiter} the shock is considerably more oblique ($\Theta_{Bn}=44\degr$). The magnetic field jump ratio is very similar at both locations. \textit{Wind} analysis further shows that shock is slow (speed 356 km/s) and relatively weak (Mach number 2.0).

\begin{table}[htbp]
\centering
	\caption{Shock Parameters at \textit{Wind} and \textit{Solar Orbiter} positions}
\renewcommand{\arraystretch}{1.2}
\begin{tabular}{c c c}\hline\hline
 & Wind & SolO \\
 $\bf{\hat n}$ [in RTN] & $(0.93, -0.02, 0.38)$ & $(0.97, -0.23, -0.05)$ \\
 $\Theta_{Bn}$ [$^\circ$]& $72$ & $44$ \\
 $V_{sh}$ [km/s]& $356$& $-$ \\
 $V_{u}$ [km/s]& $301$& $-$ \\
 $B_u$ [nT]& $2.93$ & $2.8$ \\
 $V_d/V_u$& $1.15$ & $-$ \\
 $B_d/B_u$& $2.16$ & $2.23$ \\
 $|\Delta \bf{V} \cdot \bf{\hat n}|$ [km/s]& $49.5$ & $-$ \\
 $n_{pd}/n_{pu}$ & $2.53$ & $-$ \\
 $T_{pd}/T_{pu}$ & $3.09$ & $-$ \\
 $V_{Au}$ [km/s]& $36.7$ & $-$ \\
 $V_{fu}$ [km/s]& $41.0$ & $-$ \\
 $\beta_{pu}$ & $0.3$ & $-$ \\
 $M_{fu}$ & $2.0$ & $-$ \\ \hline
\hline
\end{tabular}
\label{table:TS}
\end{table}

\begin{figure}[htbp]
\centering
\includegraphics[width=1.0\linewidth]{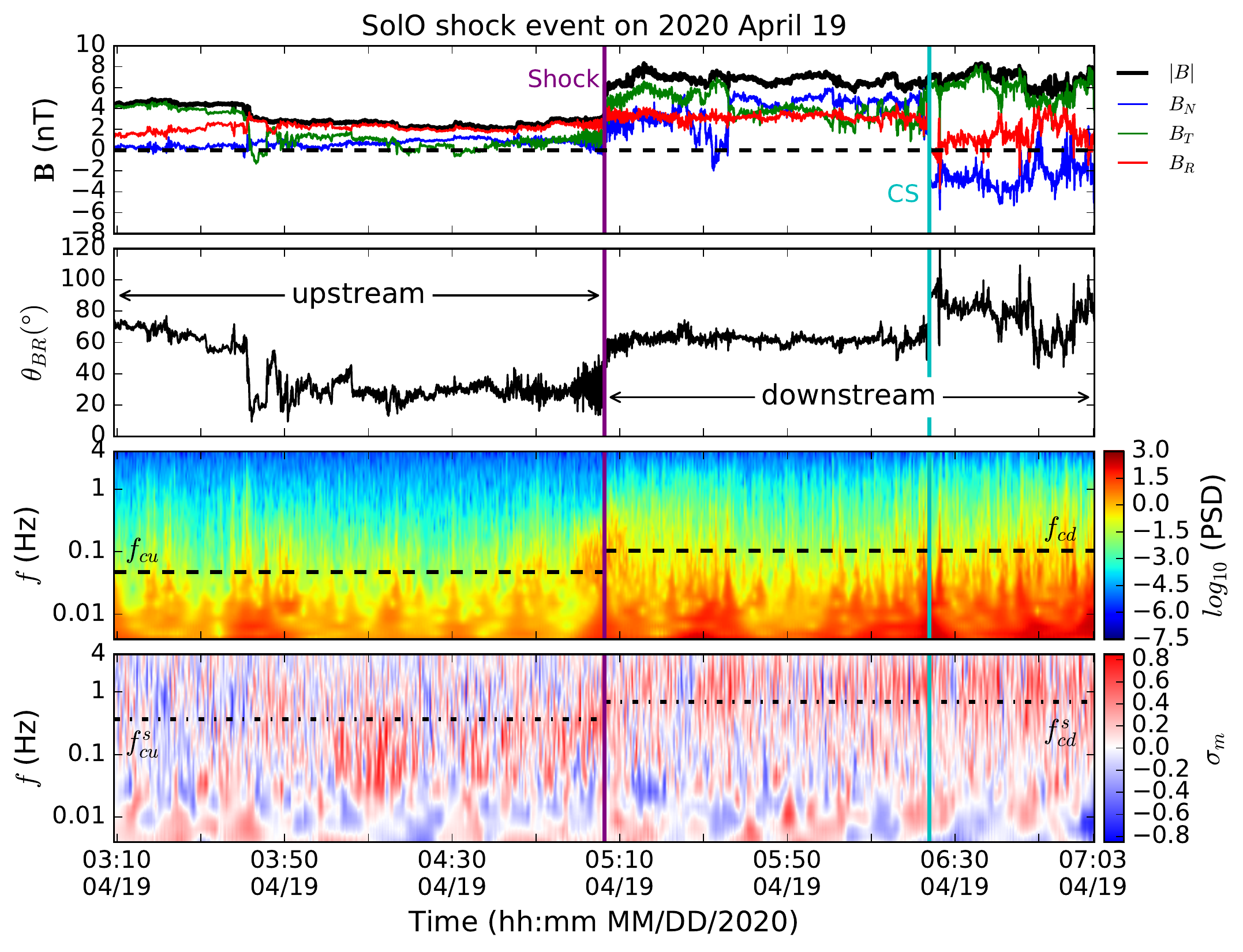} 
	\caption{\textit{Solar Orbiter} observation near the shock. The panels from top to bottom show the magnetic field components and magnitude, the angle $\theta_\mathrm{BR}$ between the magnetic field and radial direction , the total magnetic field power spectral density (PSD) and the normalized magnetic helicity $\sigma_m$ from the Morlet wavelet analysis. The cyan vertical line in each panel represents the current sheet crossing. The horizontal dashed lines in the third panel identify the proton cyclotron frequency in the plasma frame upstream $f_{cu}$ and downstream $f_\mathrm{cd}$. The dashed dotted lines in the fourth panel shows the equivalent frequency in the spacecraft frame $f_\mathrm{cu}^s$ and $f_\mathrm{cd}^s$ based on \textit{Wind}'s plasma measurements. }\label{fig:solo-shock}
\end{figure}

To study the wave activity upstream and downstream of the shock,
we now consider an interval starting 2 hours prior to the shock and ending 2 hours after the shock passage.
For both \textit{Wind} and \textit{Solar Orbiter}, the two-hour downstream interval is within the ICME sheath.
Figure \ref{fig:solo-shock} shows \textit{Solar Orbiter}'s observation of the magnetic field in this 4-hour interval. 
The ICME sheath includes a small magnetic flux rope just after 07:03 UT ahead of the ICME ejecta, which is not the focus in this study.
The magnetic field data used here has a resolution of $\sim$0.125 seconds. \textit{Solar Orbiter} is mostly in the outward magnetic sector ($B_R$ > 0) during this period.
The shock jump in the magnetic field magnitude is clearly seen in the top panel. There is a current sheet crossing around 06:23:54 UT, where the magnetic field $B_N$ and $B_R$ components change direction and the magnetic field magnitude $|B|$ has a slight drop from $\sim$ 7 nT to 5.5 nT. Visual inspection of the magnetic field time-series shows that the downstream magnetic field exhibits a higher level of fluctuations compared with the upstream.  
 The upstream magnetic field near the shock is more radially aligned compared to the downstream magnetic field, as indicated by the angle $\theta_{\mathrm{BR}}$. 
In the bottom two panels, we show the total magnetic field fluctuation power spectral density (PSD) and the normalized magnetic helicity $\sigma_\mathrm{m}$ from the wavelet analysis.
The PSD shows that the downstream fluctuating power is higher compared to that upstream at a fixed frequency.
As a reference, the proton cyclotron frequencies in the plasma frame are plotted in the third panel both upstream ($f_\mathrm{c,u}$) and downstream ($f_\mathrm{c,d}$), and the equivalent frequency in the spacecraft frame $f_\mathrm{c,u}^s = f_\mathrm{c,u} \cdot V_\mathrm{sw,u}/V_\mathrm{A,u}$ and $f_\mathrm{c,d}^s = f_\mathrm{c,d} \cdot V_\mathrm{sw,d}/V_\mathrm{A,d}$ are shown in the bottom panel. Here, $V_\mathrm{sw}$ and $V_\mathrm{A}$ are the solar wind speed and Alfv\'en speed estimated by \textit{Wind}'s plasma measurements.      
The spectrogram of the normalized magnetic helicity shows a dominance of positive and relatively large $\sigma_\mathrm{m}$ around 0.1 Hz within an hour prior to the shock crossing ($\sim$ 04:00--05:00).
This indicates the existence of right-hand polarized waves in the outward magnetic sector. After crossing the shock, these positive and large $\sigma_\mathrm{m}$ values are observed and are prevalent at a frequency slightly higher than $f^s_\mathrm{cd}$. The wave frequency increases by a factor of $\sim$10 with the shock crossing. 
This can be understood from the flow velocity $V_{sw}$ increases in the downstream, causing the frequency $f_{sc}$ increase due to the Doppler effect. In addition, compression causes the wave number $k$ to increase (or wavelength $\lambda$ to decrease) across the shock \citep{zank2021}, which also increases the observed frequency by Taylor's hypothesis ($f_{SC}=V_{sw}/\lambda$, $\lambda=2\pi/k$).

\begin{figure}[htbp]
\centering
\includegraphics[width=1.0\linewidth]{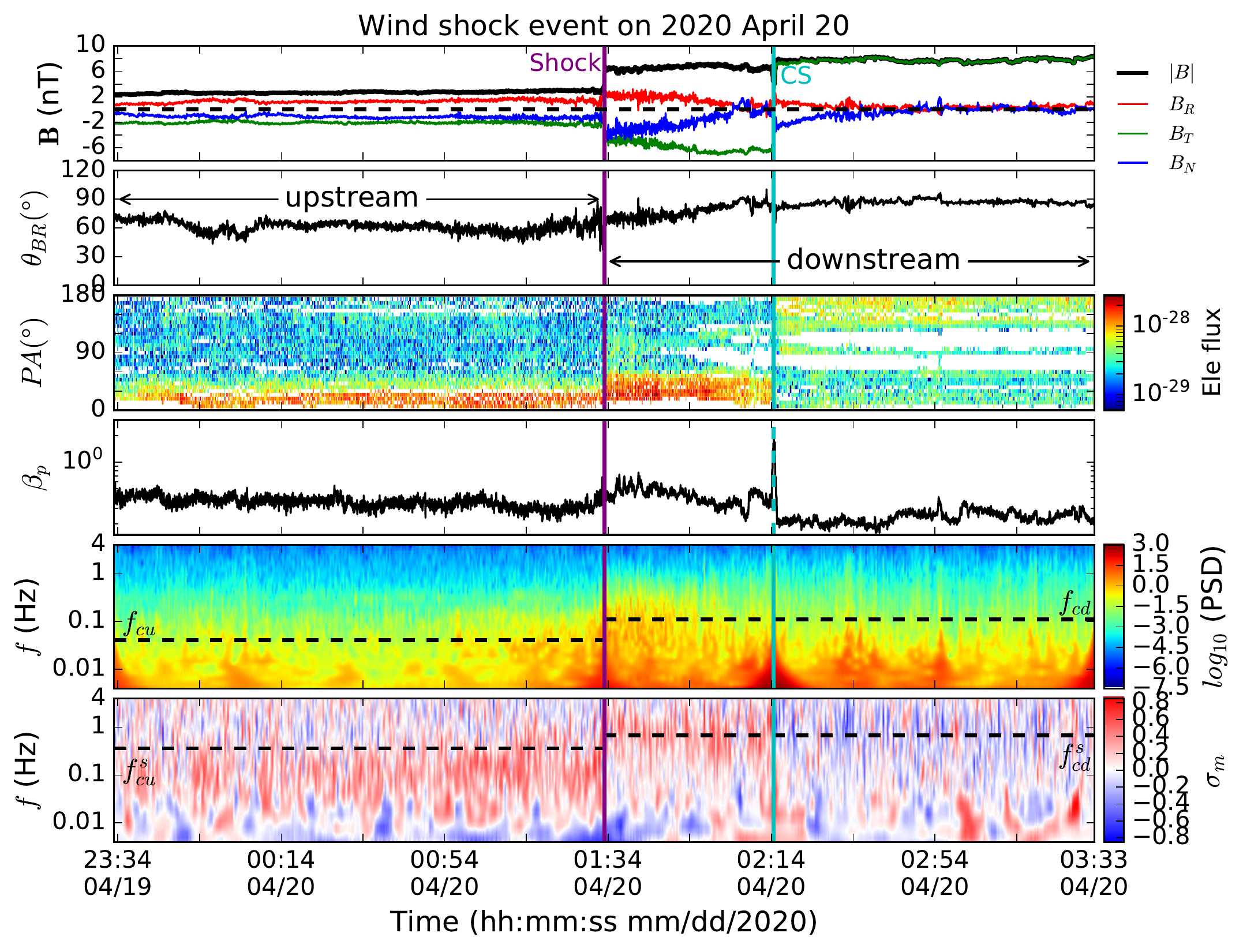} 
\caption{In the same format as Fig.\ \ref{fig:solo-shock}, but for \textit{Wind} observation near the shock. The third panel shows the pitch angle distribution of 97.37 eV electrons, and the fourth panel shows the proton plasma beta during this period. The cyan vertical lines represent the current sheet crossing.}\label{fig:wind-shock}
\end{figure}

The same analysis is done for the \textit{Wind} data, as shown in Figure \ref{fig:wind-shock}. Here, magnetic field data with a resolution of 0.092 seconds are used. We also show the pitch angle distribution (PAD) of 97.37 eV electrons and the proton plasma beta $\beta_p$ to characterize a strong current sheet (SCS) crossing during this period. The SCS is identified by the directional change of the magnetic field, the decrease of the magnetic field magnitude $|B|$, and a sharp increase in the proton plasma beta $\beta_p$. In the electron PAD panel, the unidirectional electron beam initially is aligned with 0$\degr$ pitch angle and then switches to 180$\degr$. Based on the multiple proton plasma beta jumps in the ICME-sheath observed by \textit{Wind}, it can be related to the heliospheric current sheet (HCS) since the flow is slow and may originate from the streamer belt, in which the HCS is often embedded. 
Unlike \textit{Solar Orbiter}, the \textit{Wind} magnetic field appears to be quasi-perpendicular to the radial direction both upstream and downstream of the shock as shown in the $\theta_{\mathrm{BR}}$ panel. The PSD panel shows that similar to \textit{Solar Orbiter}, the magnetic fluctuation power increased in the downstream region. Compared to the PSD measured by \textit{Solar Orbiter}, the magnetic fluctuation power observed by \textit{Wind} is smaller, illustrating that the magnetic fluctuation power decreases as distance increases \citep[e.g.,][]{telloni2015}.  
The characteristics of magnetic helicity are similar to those in Figure \ref{fig:solo-shock}. In the upstream region, an enhanced magnetic helicity ($>$0) is also observed near 0.1 Hz. However, this phenomenon seems to be prevalent throughout the upstream 2-hour interval, which is different from that observed by \textit{Solar Orbiter} (within one hour).
After crossing the shock, a large and positive magnetic helicity is observed near the downstream proton cyclotron frequency $f_\mathrm{cd}^s$ but lasts only about 40 minutes. The magnitude of $\sigma_m$ is slightly smaller than upstream.
After the SCS crossing, no significant positive enhancement is shown in the spectrogram of $\sigma_m$. 

\begin{figure}[htbp]
\centering
\includegraphics[width=0.8\linewidth]{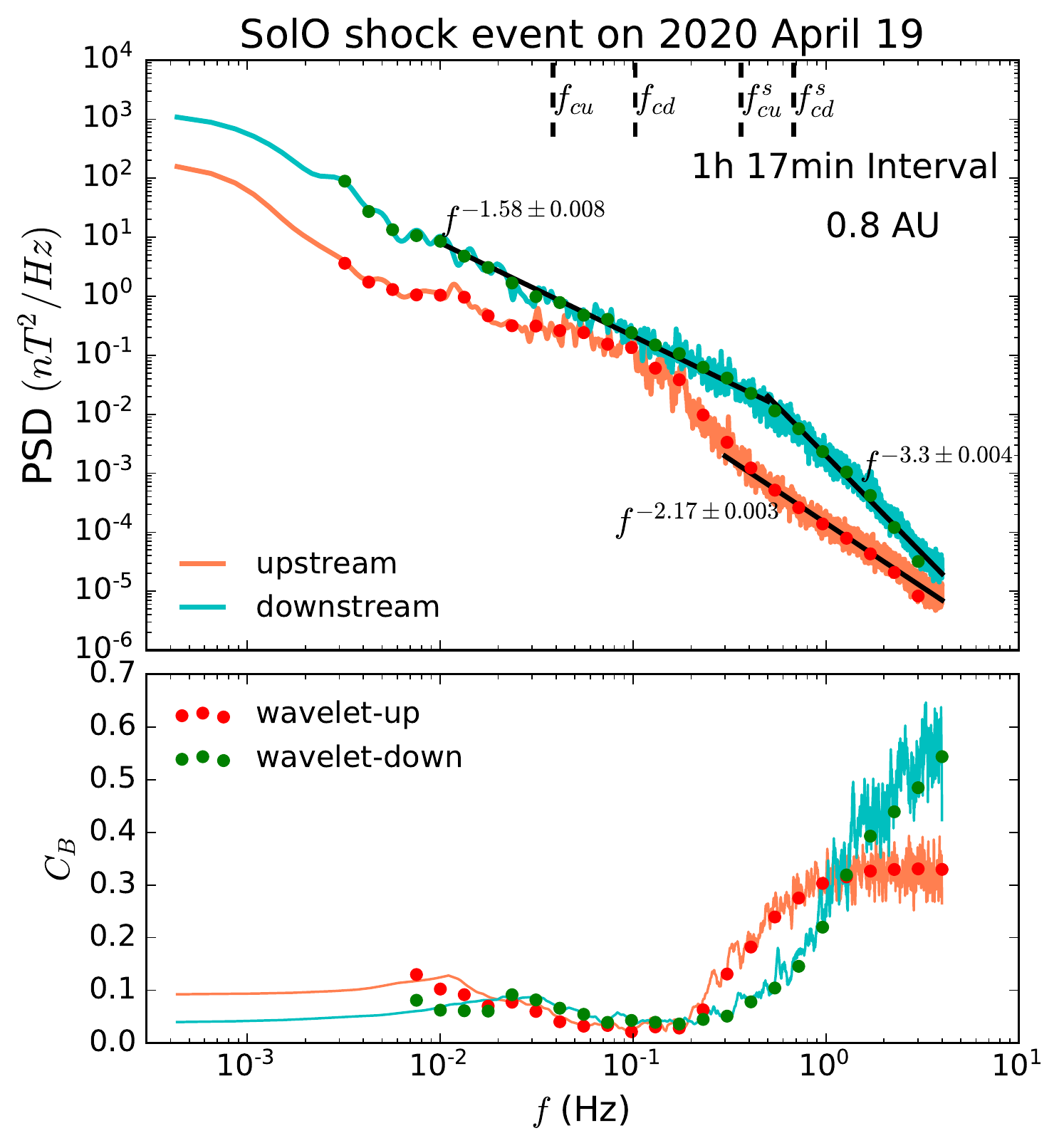} 
\caption{\textit{Solar Orbiter}'s total magnetic field power spectra density (top panel) and magnetic comressibility $C_B$ (bottom panel) as a function of frequency upstream and downstream of the shock. The red and turquoise lines represent the upstream and downstream Fourier spectra using the Blackman-Tukey method. The red and green dots represent the corresponding Morlet wavelet results. Each spectrum is calculated in a 77-minute interval upstream and downstream, respectively. Black straight lines in the top panel show power-law fits, and vertical dashed lines indicate the upstream  and downstream proton cyclotron frequency in the plasma frame $f_\mathrm{cu}$, $f_\mathrm{cd}$ and in the spacecraft frame$f_\mathrm{cu}^s$, $f_\mathrm{cd}^s$.}\label{fig:solo-spectra}
\end{figure}

In Figure \ref{fig:solo-spectra}, we show the frequency dependent magnetic fluctuation trace PSD and magnetic compressibility $C_\mathrm{B}$ upstream and downstream of the shock observed by \textit{Solar Orbiter}. We use both the standard Fourier method and the wavelet technique.
To avoid the possible effects of the current sheet, the downstream spectrum is calculated within a 77-minute interval (05:06:18--06:23:18 UT), corresponding to the region between the ``shock'' and ``CS'' shown in Figure \ref{fig:solo-shock}. The upstream spectrum is computed at the same interval length, i.e., 03:49:18--05:06:18 UT.  
The Fourier spectrum is calculated using the Blackman-Tukey method, i.e., the Fourier transform of the correlation function
$ \mathrm{PSD} = \mathcal{F}[R(\tau)] $. The upstream spectrum is plotted in red and downstream in turquoise.
The solid lines are the Fourier spectra and the dots are the wavelet spectra, which are consistent with each other.
Clearly, the trace power of the magnetic field fluctuations is enhanced downstream.
The amplitude of the magnetic fluctuations $|\delta \vec{B}|$, i.e., the integral of the Fourier PSD, is $\sim$0.6 nT upstream and $\sim$1.68 nT downstream. 
The upstream spectrum shows a bump in the frequency range between $f_\mathrm{cu}$ and $f_\mathrm{cu}^s$. The averaged $\theta_\mathrm{BR}$ during the upstream interval is around 30$\degr$ as shown in Figure \ref{fig:solo-shock}. The significant enhancement of the upstream magnetic fluctuation power at around 0.1 Hz indicates the presence of quasi-parallel propagating waves.
The downstream inertial-range spectrum follows a power-law shape and is close to the Kolmogorov $-5/3$ spectrum.
At higher frequencies, both upstream and downstream spectra steepen. In the frequency range of 0.3 to 4 Hz, the upstream spectrum has a slope of $f^{-2.17}$. The downstream break frequency is around 0.5 Hz. The spectral break frequency can be estimated by $f_b = V_{sw}/2\pi/(d_i+\rho_i)$ with $d_i$ the proton inertial length and $\rho_i$ thermal proton gyroradius \citep{duan2018}. The proton speed, density and temperature needed here are obtained from \textit{Wind} measurements. After the break frequency, the downstream spectrum behaves like $f^{-3.3}$.
The magnetic compressibility $C_B$ is defined as the ratio between the power in the magnetic field magnitude fluctuations and the power in total fluctuation ($P_\mathrm{|B|}/P_\mathrm{Tr}$) \citep{bavassano1982}.
As the solar wind turbulence is typically incompressible, $C_B$ is usually smaller than 0.1 in the inertial range. As it approaches the kinetic range with high frequencies, the compressibility increases obviously as shown in the figure. The upstream compressibility near 0.1 Hz is slightly smaller than downstream $C_\mathrm{B}$. In the frequency range 0.02--0.2 Hz, the upstream $C_B$ is around 0.04 and downstream $C_B$ is around 0.06. The downstream compressibility shows a significant increase after its spectral break frequency and exceeds the upstream $C_B$ at frequencies above 1 Hz.  

\begin{figure}[htbp]
\centering
\includegraphics[width=0.5\linewidth]{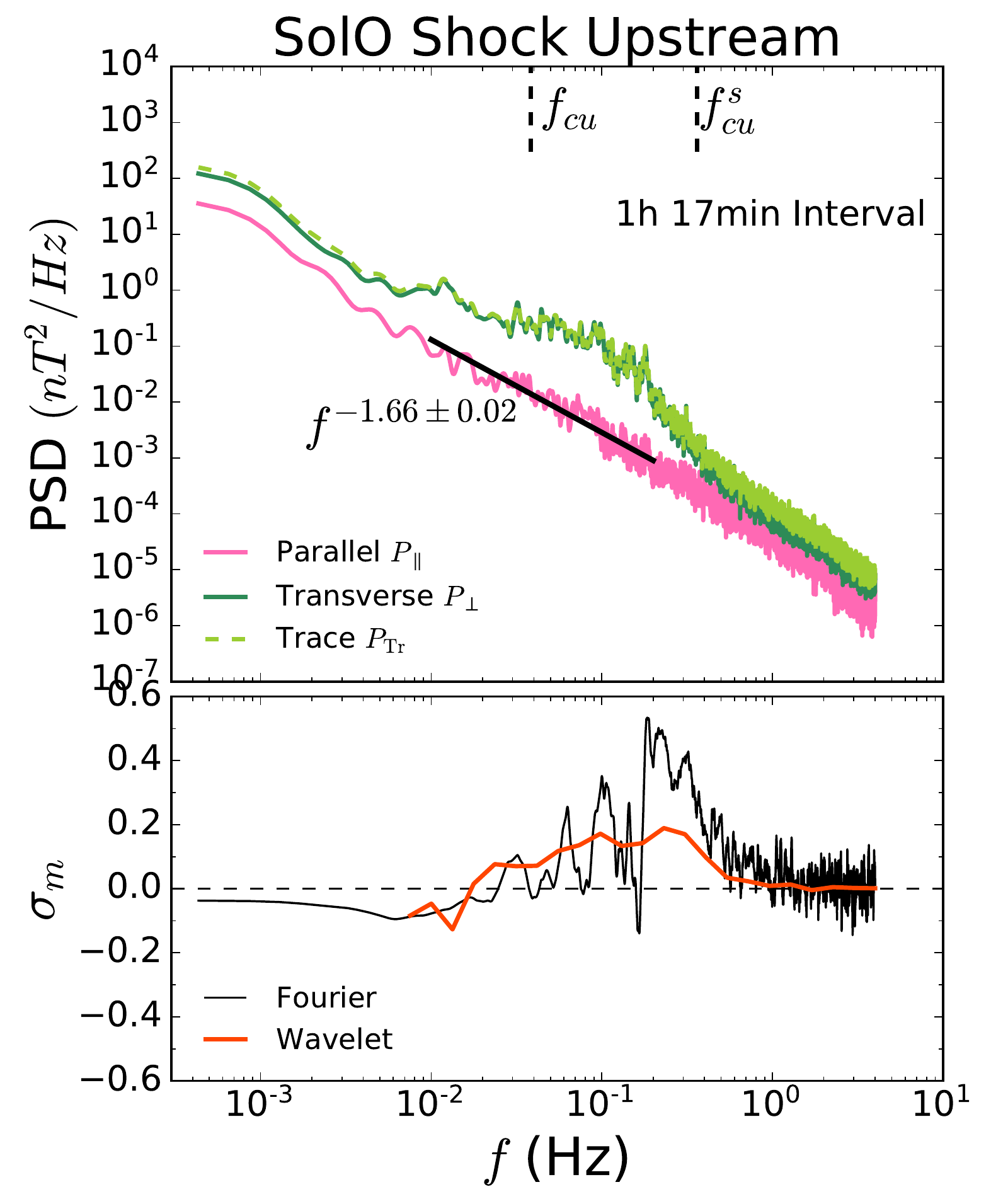}%
\includegraphics[width=0.5\linewidth]{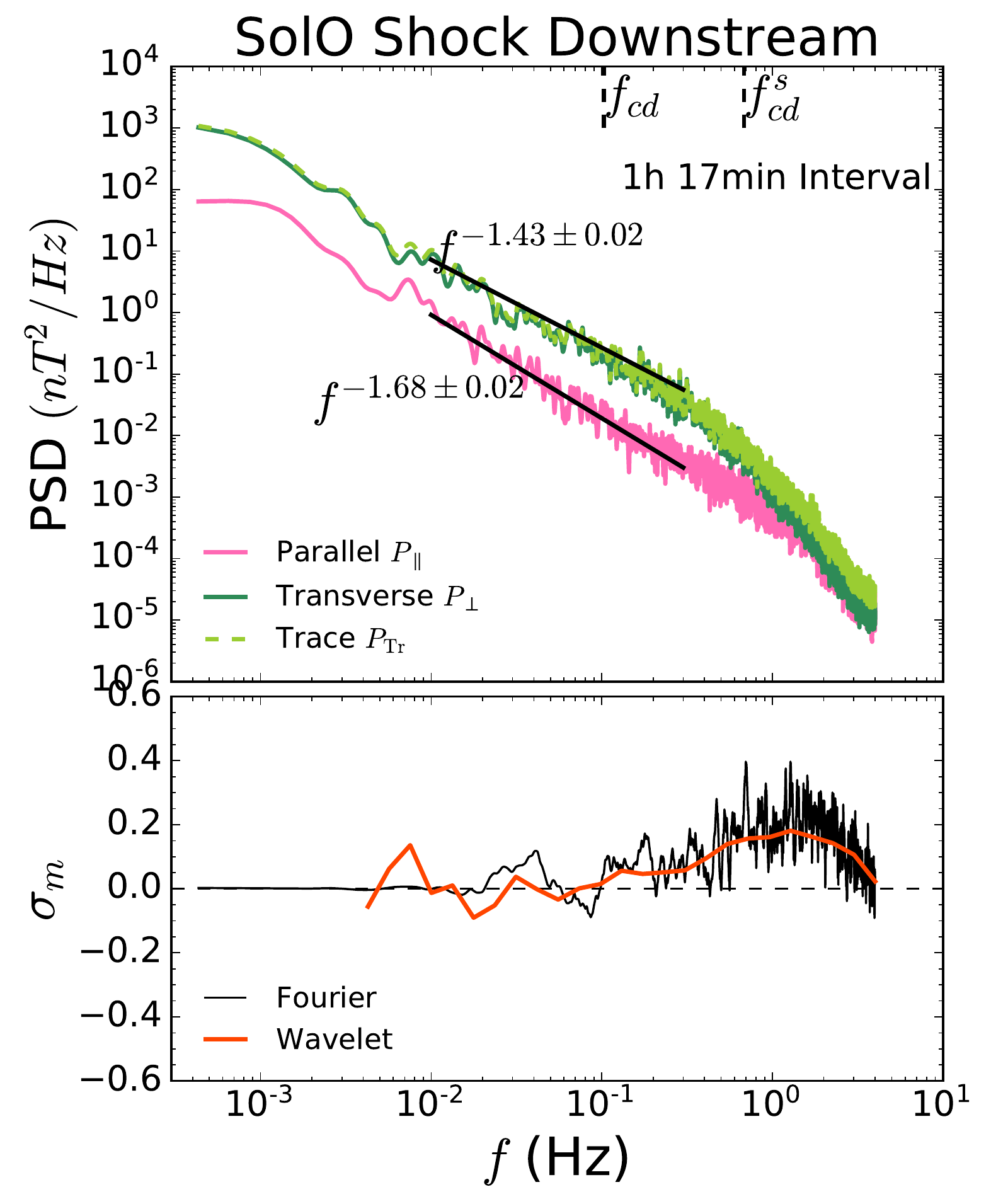} 
\caption{The top panels show \textit{Solar Orbiter}'s spectra for total magnetic field fluctuations (limegreen dashed lines) separated into compressible parallel (pink solid lines) and incompressible transverse fluctuations (green solid lines) in the upstream (left panels) and downstream (right panels) regions, respectively. The upstream and downstream proton cyclotron frequency in plasma frame $f_\mathrm{cu}$ and $f_\mathrm{cd}$ and the equivalent frequency in the spacecraft frame $f_\mathrm{cu}^s$ and $f_\mathrm{cd}^s$ are shown as well. Power-law fitting is performed in the frequency range [0.01, 0.2] Hz. The bottom panels show the spectra of the normalized magnetic helicity spectra $\sigma_m$ calculated using both the Blackman-Tukey (black lines) and Morlet wavelet (red lines) methods.}\label{fig:solo-para-perp}
\end{figure}

The magnetic compressibility can also be represented by the ratio between fluctuations parallel and perpendicular to the mean magnetic field $P_\mathrm{\parallel}/P_\mathrm{\perp}$, as shown in Figure \ref{fig:solo-para-perp}.
In the top panels, the parallel and transverse spectra are plotted in pink and green, respectively, and the sum of the two (trace spectra) in limegreen. The trace spectra are the same as in Figure \ref{fig:solo-spectra}.
Transverse spectra dominate both upstream and downstream, indicating the dominance of nearly incompressible fluctuations \citep{Zank2017}.
Another notable feature is that the bump near 0.1 Hz in the upstream PSD and the spectral break of the downstream PSD at around 0.5 Hz are both caused by the transverse fluctuations. The ratio of parallel fluctuation power to the perpendicular fluctuation power $P_\mathrm{\parallel}/P_\mathrm{\perp}$ (not shown here) upstream and downstream is consistent with their respective magnetic compressibility $C_B=P_\mathrm{|B|}/P_\mathrm{Tr}$ shown in Figure \ref{fig:solo-spectra}.  
The bottom panels of Figure \ref{fig:solo-para-perp} show the Fourier and time-averaged wavelet spectra of the normalized magnetic helicity $\sigma_\mathrm{m}$ as a function of the frequency.
Enhanced magnetic helicity is another signature of wave activities as the solar wind is usually in a state with $\sigma_m \simeq 0$ \citep[e.g.,][]{vasquez2018}.  
Both upstream and downstream $\sigma_m$ spectra exhibit a positive bump, suggesting the existence of right-hand polarized wave modes in the spacecraft frame.
The upstream spectral bump and the enhancement of $\sigma_m$ appear in a wide and low frequency range, i.e., 0.01--0.4 Hz, while the downstream $\sigma_m$ increases at relatively high frequencies near $f^s_\mathrm{cd}$ and peaks at around 1 Hz. 

\begin{figure}[htbp]
\centering
\includegraphics[width=0.8\linewidth]{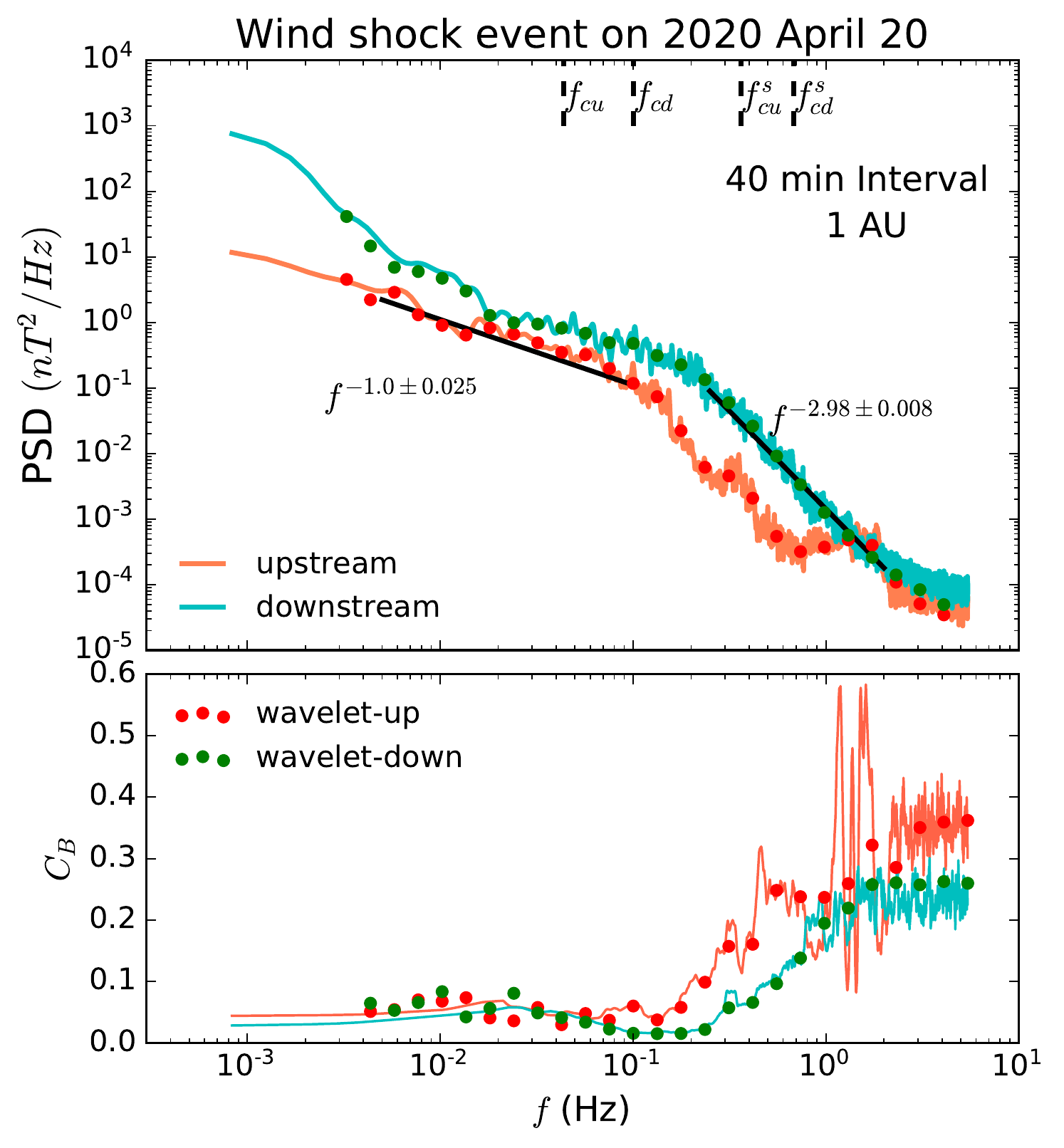} 
\caption{In the same format as Fig.\ \ref{fig:solo-spectra}, but for \textit{Wind} observations upstream and downstream of the shock. Vertical dashed lines in the top panel indicate the proton cyclotron frequency in the plasma frame upstream $f_\mathrm{cu}$ and downstream $f_\mathrm{cd}$ and in the spacecraft frame upstream $f^s_\mathrm{cu}$ and downstream $f^s_\mathrm{cu}$, respectively. Black solid lines in the top panel show the power-law fits. The power spectra and compressibility $C_B$ are calculated within a 40-minute interval. }\label{fig:wind-spectra}
\end{figure}

\begin{figure}[htbp]
\centering
\includegraphics[width=0.5\linewidth]{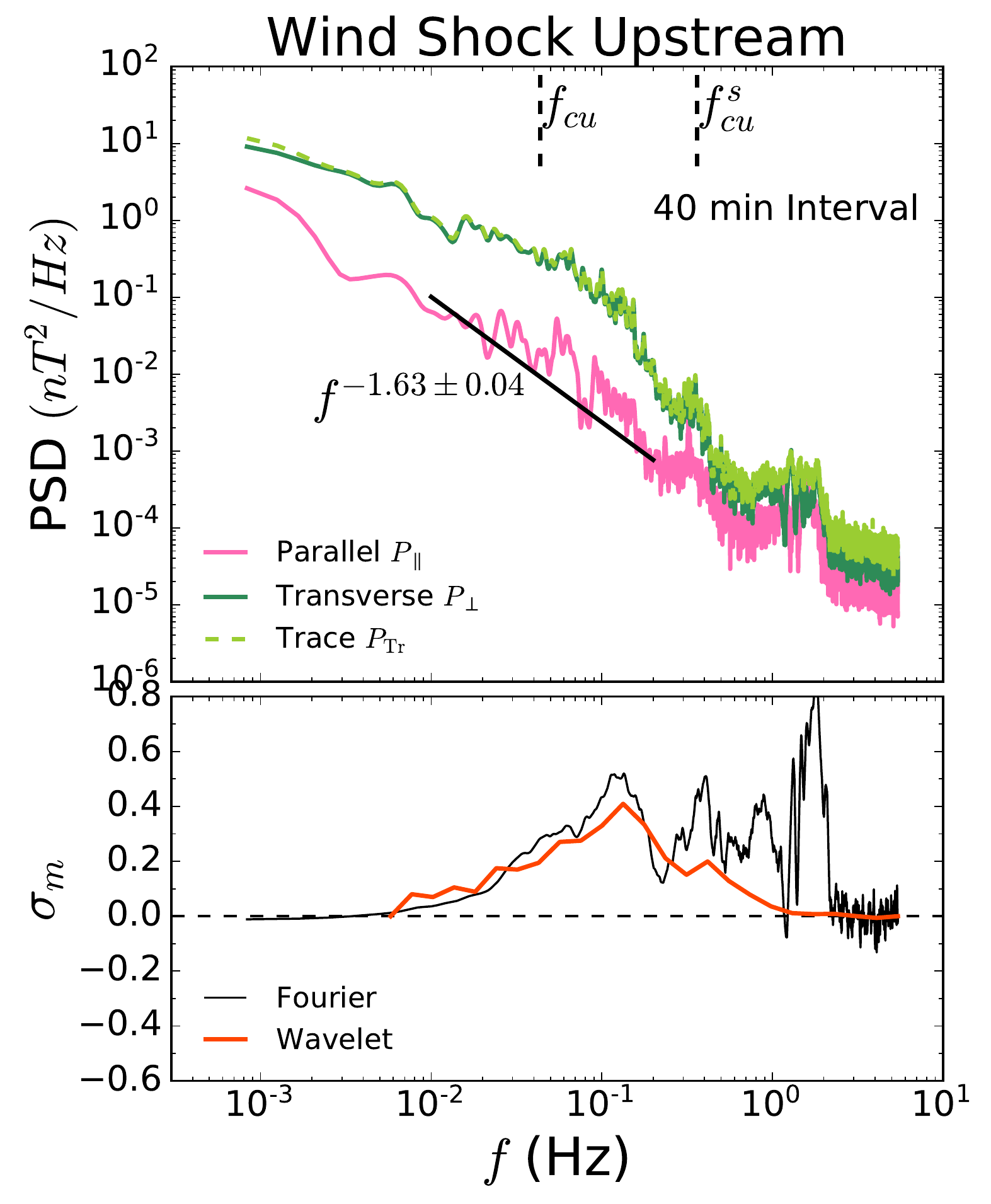}%
\includegraphics[width=0.5\linewidth]{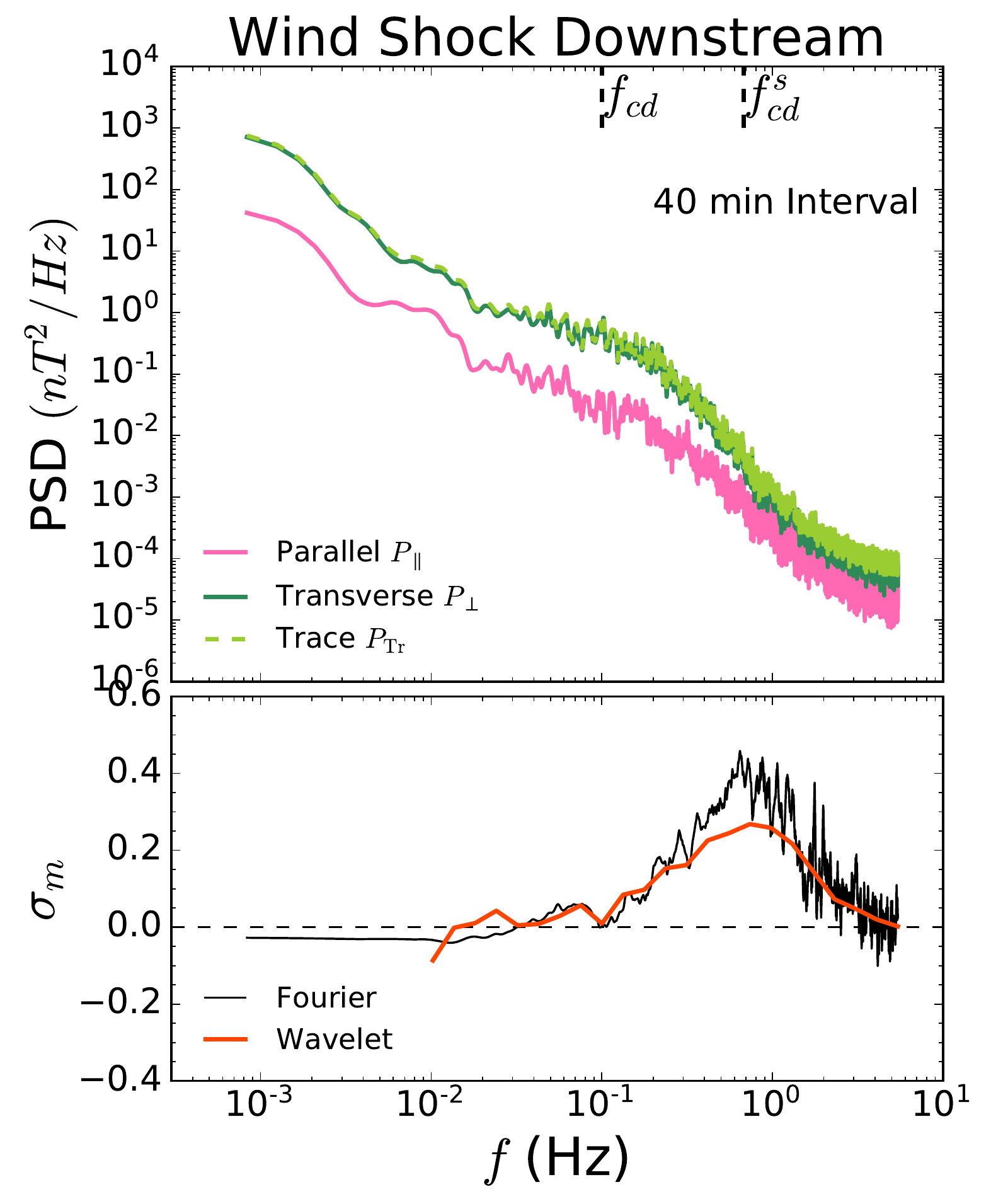}%
\caption{In the same format as Fig.\ \ref{fig:solo-para-perp}, but for \textit{Wind} observations. Top panels show the upstream (left) and downstream (right) power spectra of the total magnetic fluctuations (limegreen dashed lines), compressible fluctuations (pink solid lines), and incompressible traverse fluctuations (green lines). Bottom panels show the spectra of normalized magnetic helicity $\sigma_m$ obtained from both Fourier and Wavelet methods. }\label{fig:wind-para-perp}
\end{figure}

The same spectral analysis for \textit{Wind} data near the shock is presented in Figures \ref{fig:wind-spectra} and \ref{fig:wind-para-perp}. To avoid the effects of the strong current sheet observed by \textit{Wind} shown in Figure \ref{fig:wind-shock},
each power spectrum here is calculated within a 40-minute interval prior to the shock front (upstream) and after the shock passage (downstream).
\begin{table}[htbp]
\centering
\caption{Magnetic field and solar wind parameters upstream and downstream of the shock observed by \textit{Wind}}
\renewcommand{\arraystretch}{1.2}
\begin{tabular}{c c c}\hline\hline
 & Upstream & Downstream  \\
 $|B|$ [nT]               & $2.86$    & $6.59$ \\
 $V_{sw}$ [km/s]          & $299$     & $345$  \\
 $n_{p}$ [$cm^{-3}$]      & $3.0$     & $8.0$ \\
 $T_{p}$ [K]              & $25056$   & $65746$ \\
 $V_A$ [km/s]             & $36$      & $51$ \\
 $\beta_p$                & $0.32$    & $0.42$ \\
 $f_\mathrm{c}$[Hz]       & $0.044$   & $0.1$ \\
 $f^s_\mathrm{c}$[Hz]     & $0.36$    & $0.68$ \\
 $d_\mathrm{i}$[km]       & $131$     & $81$ \\
 $\rho_\mathrm{i}$[km]    & $52$     & $37$ \\
 $|\delta \vec{B}|$ [nT]        & $0.4$    & $1.57$ \\ \hline
\hline
\end{tabular}
\label{table:up}
\end{table}
Table \ref{table:up} lists \textit{Wind} measurements of the magnetic field and flow plasma parameters during this period. All the parameters are the mean of the 40-minute interval. The magnetic fluctuation amplitude $|\delta \vec{B}|$ increases about 4 times downstream, while it increases about three times at the downstream observed by \textit{Solar Orbiter}. 

Similar to the \textit{Solar Orbiter} results, we find an amplification in the magnetic field PSD downstream of the shock. However, due to the wave activity both upstream and downstream, the amplification is not a constant shift. Power-law fitting is performed on the upstream spectrum in the frequency range [0.005, 0.1] Hz and a flatter spectrum with $f^{-1.0}$ is obtained. After about 0.1 Hz, the upstream spectrum starts to steepen. There are two other enhancements present at high frequencies in the upstream spectrum, but may be related to instrument noise. 
The downstream power spectrum also deviates from a single power-law spectrum. The spectrum starts to steepen after about 0.2 Hz. The magnetic compressibility $C_B$ at high frequencies ($>0.1$ Hz) is larger in the upstream region compared to the downstream, but little difference is present at low frequencies.

Figure \ref{fig:wind-para-perp} is in the same format as Figure \ref{fig:solo-para-perp}. The top panels show the power spectra of the total magnetic fluctuations $P_\mathrm{Tr}$, compressible fluctuations $P_{\parallel}$, and transverse fluctuations $P_{\perp}$. Again, the magnetic field fluctuation power is dominated by the incompressible traverse fluctuations. The ratios $P_{\parallel}/P_{\perp}$ upstream and downstream are consistent with the magnetic compressibility obtained by $C_\mathrm{B} = P_{|B|}/P_\mathrm{Tr}$ shown in Figure \ref{fig:wind-spectra}, which is also found in \textit{Solar Orbiter}'s results. The upstream wave activity near 0.1 Hz is mostly in the traverse fluctuations. The downstream transverse spectrum shows a bump in the frequency range around 0.2 Hz. The wave activity is also reflected in the spectrum of the normalized magnetic helicity $\sigma_m$, shown in the bottom panel. Both upstream and downstream wave activities show a positive enhancement of $\sigma_m$, indicating also a right-handed polarization. The upstream $\sigma_m$ wavelet spectrum peaks around 0.1 Hz and the downstream $\sigma_m$ peaks around 0.7 Hz (close to $f^s_\mathrm{cd}$). The downstream wave frequency is clearly larger than that upstream. The upstream wave signature falls over a relatively wide and low frequency range, resulting in a flat $f^{-1}$ spectrum at frequencies less than 0.1 Hz. This indicates the low-frequency (0.01--0.1 Hz) right-hand polarized waves \citep{he2019}, which are often observed in planetary foreshocks as ultra-low-frequency (ULF) waves \citep[e.g.,][]{greenstadt1995ulf, narita2003}. In contrast, the downstream wave activity appears near $f^s_\mathrm{cd}$ and is right-hand polarized, which might be kinetic Alfv\'en waves observed in the solar wind \citep[e.g.,][]{he2011oblique, podesta2013evidence, telloni2020wave}. 

\section{Upstream and downstream waves}

In this section, we further analyze the nature of the waves observed in this shock event.
The wave propagation angle relative to the mean magnetic field is crucial to the analysis.
To find the angle, we calculate the local mean magnetic field based on the envelope of the wavelet function \eqref{eq:wavelet},
\begin{equation}
  \bar{B}(s, t_n) = \sum_m B(t_m) \exp\left[-\frac{(t_n - t_m)^2}{2 s^2}\right],
\end{equation}
which depends on both the scale $s$ and time $t$.  
We can then calculate the angle between the local mean magnetic field $B_\mathrm{0l}$ and the radial direction $\theta_{\mathrm{B_{0l}R}}(s, t)$, which also depends on scale and time. The scale $s$ is related to the time period $p$ through $s \sim p/1.03$ for the Morlet wavelet transform.
By Taylor's hypothesis, the observed wavevector is in the direction of the solar wind speed. For the sake of simplicity, we assume that the solar wind velocity is approximately radial.
Therefore, the angle $\theta_{\mathrm{B_{0l}R}}(s, t)$ represents the wave propagation angle relative to the local mean magnetic field.

\begin{figure}[htbp]
\centering
\includegraphics[width=0.5\linewidth]{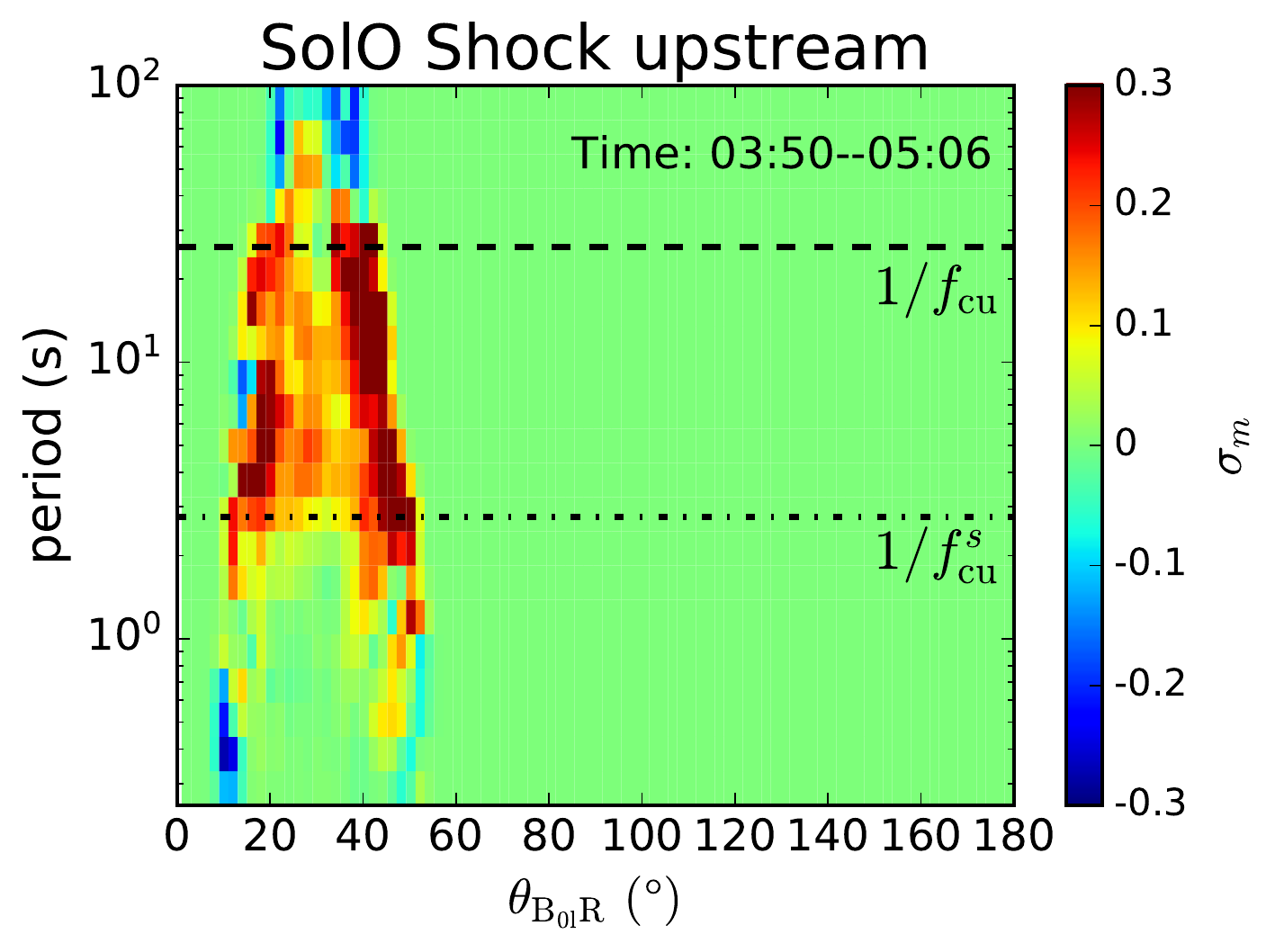}%
\includegraphics[width=0.5\linewidth]{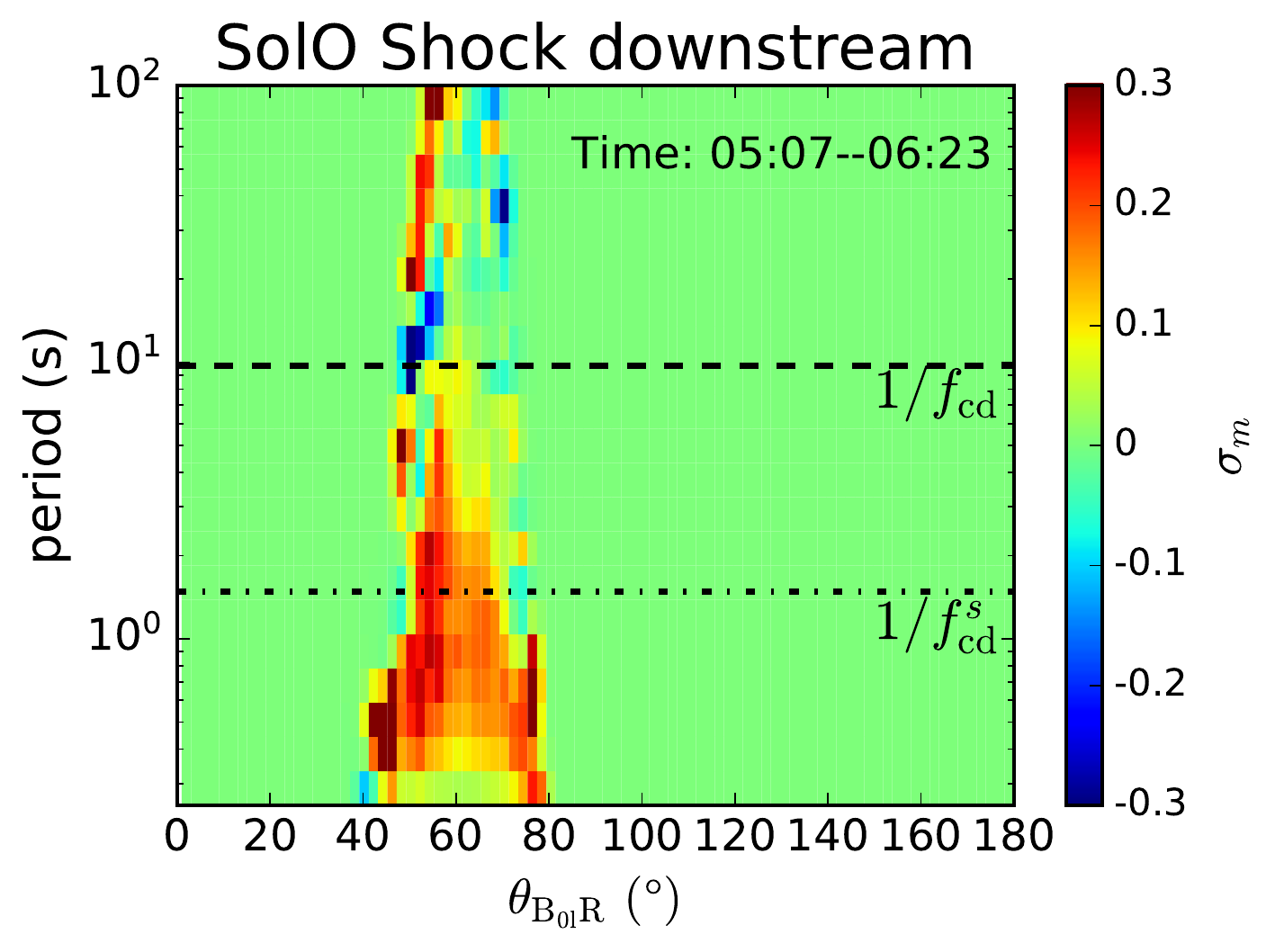}%
	\caption{$\theta_\mathrm{B_{0l}R}$ distributions of $\sigma_m$ spectra upstream (left) and downstream (right) of the shock observed by \textit{Solar Orbiter}. $\theta_\mathrm{B_{0l}R}$ is the angle between the scale-dependent local mean magnetic field $B_\mathrm{0l}$ and the radial direction. The inverse of the period gives the
corresponding frequency. The horizontal dashed lines and dashed dotted lines show the periods corresponding to the proton cyclotron frequency in the plasma frame and in the spacecraft frame, respectively.}\label{fig:solo-theta}
\end{figure}

A histogram of the propagation angle $\theta_{\mathrm{B_{0l}R}}$ can be constructed as shown in Figure \ref{fig:solo-theta}, which illustrates the likelihood of different wave propagation direction at each scale. Here, waves in solar wind turbulence are identified by the enhanced magnetic helicity $\sigma_m$ \citep[e.g.,][]{he2011, he2011oblique, vasquez2018, telloni2019}.
The left and right panels show the $\sigma_m$ spectra as a function of the propagation angle $\theta_{\mathrm{B_{0l}R}}$ for upstream and downstream of the shock, respectively. The time interval for calculating $\theta_{\mathrm{B_{0l}R}}$ in each region is the same as in Figure \ref{fig:solo-spectra}.  
During this period, the solar wind is in the outward magnetic sector with $B_\mathrm{R}>0$. 
The results show that the positive enhancement of the upstream $\sigma_m$ in the period range [3, 26] s or frequency range [0.04, 0.3] Hz, which corresponds to the bump shown in the bottom left panel of Figure \ref{fig:solo-para-perp}, is mainly in the quasi-parallel direction, i.e., $\theta_{\mathrm{B_{0l}R}} < 45\degr$.
On the other hand, the downstream enhanced $\sigma_m$ has a more perpendicular $\theta_{\mathrm{B_{0l}R}}$ ($\sim$60$\degr$), indicating that the waves downstream are more oblique. The downstream wave frequency is about 10 times larger than the upstream wave frequency. Due to the quasi-parallel propagating angle upstream, the positively enhanced $\sigma_m$ suggests the presence of right-hand polarized quasi-parallel propagating waves in the spacecraft frame. The frequency range of upstream waves is between $f_\mathrm{cu}$ and $f^s_\mathrm{cu}$ and the downstream right-hand polarized waves are mainly in the period less than 2 s or frequency larger than 0.5 Hz. The downstream waves can be transmitted from upstream with a higher frequency due to the shock compression and Doppler shift, but can also be locally generated downstream.  

\begin{figure}[htbp]
\centering
\includegraphics[width=0.5\linewidth]{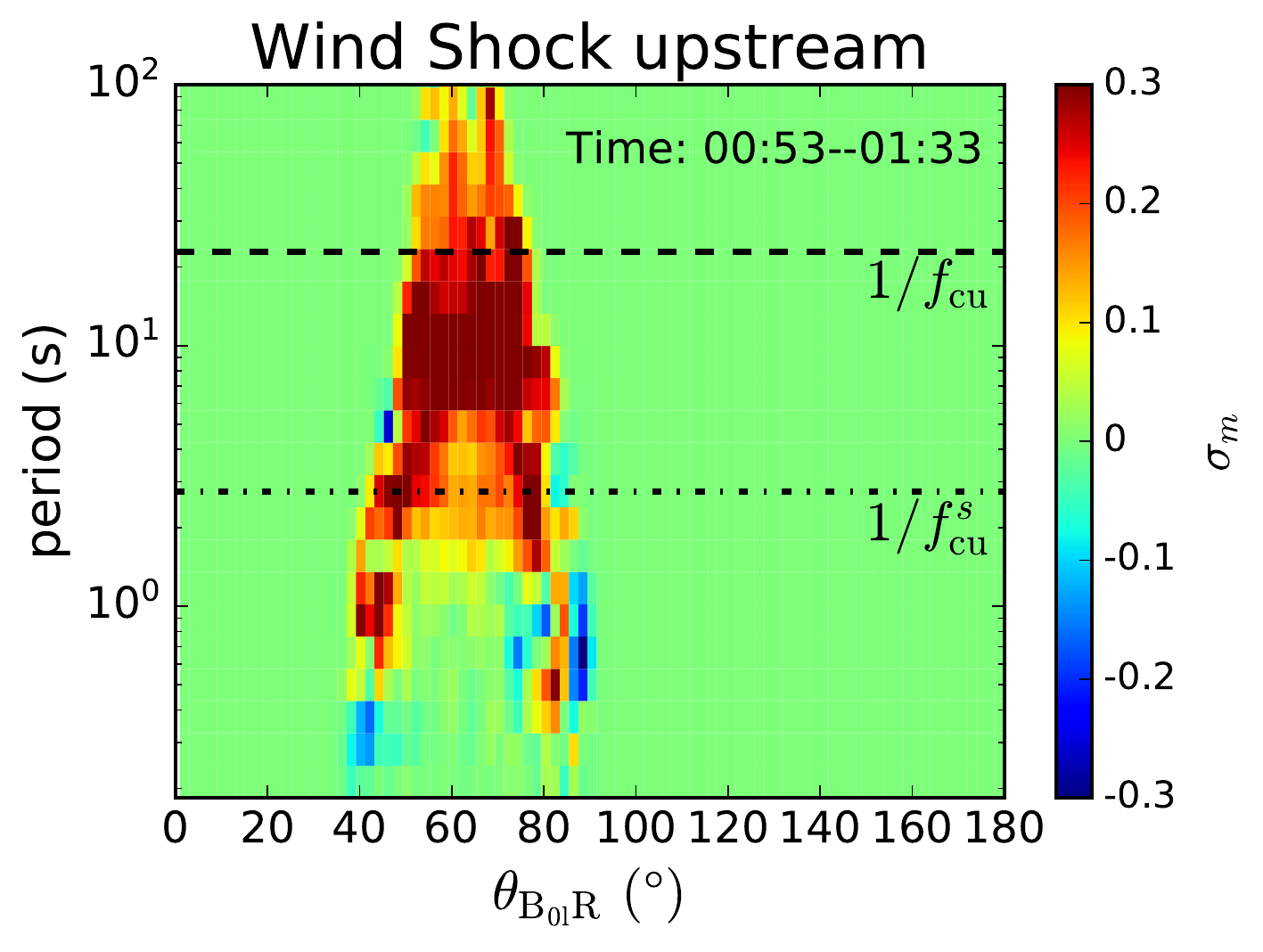}%
\includegraphics[width=0.5\linewidth]{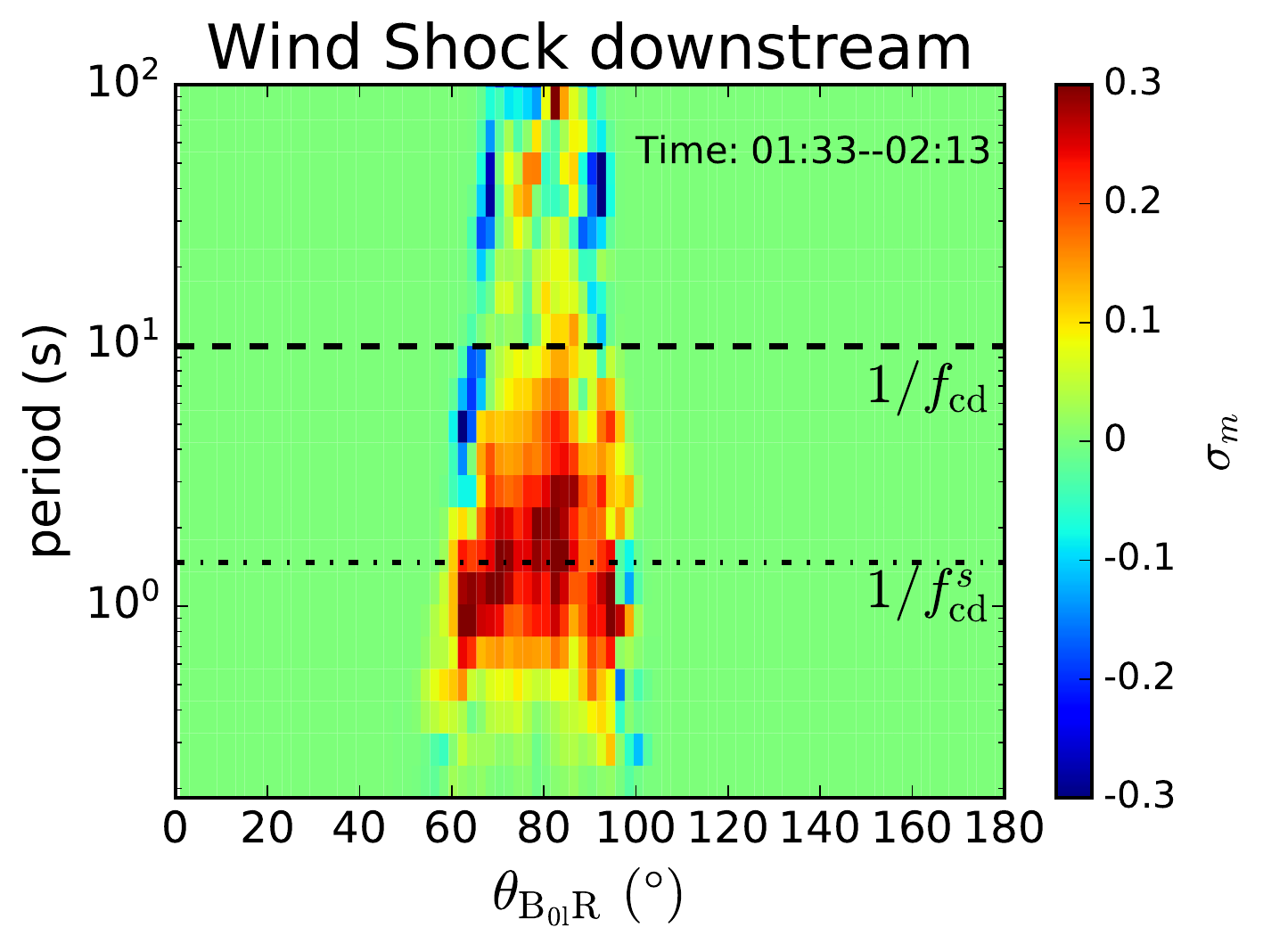}%
	\caption{In the same format as Fig.\ \ref{fig:solo-theta}, but for \textit{Wind} observation of $\theta_\mathrm{B_{0l}R}$ distribution. }\label{fig:wind-theta}
\end{figure}

Figure \ref{fig:wind-theta} shows the same analysis but for \textit{Wind} observations upstream (left) and downstream (right) of the shock. The positively enhanced $\sigma_m$ in the upstream region is observed in a wide period range between 2--100 s. The angle $\theta_\mathrm{B_{0l}R}$ is between 50$\degr$ and 80$\degr$. The wave propagation angle is more oblique to the background magnetic field compared to the \textit{Solar Orbiter} observations at 0.8 AU.
The downstream wave activity mainly occurs in the periods of 0.8--4 s, and the propagation angle $\theta_\mathrm{B_{0l}R}$ is concentrated in the range 60$\degr$--80$\degr$. As discussed above, the downstream wave frequency increases by about 10 times, but the downstream solar wind speed is only 1.15 times larger than that upstream, indicating a shorter wavelength downstream of the shock due to the compression \citep{zank2021}. The positively enhanced $\sigma_m$ observed by \textit{Wind} also suggests the existence of right-hand polarized waves both upstream and downstream. 

\begin{figure}[htbp]
\centering
\includegraphics[width=0.5\linewidth]{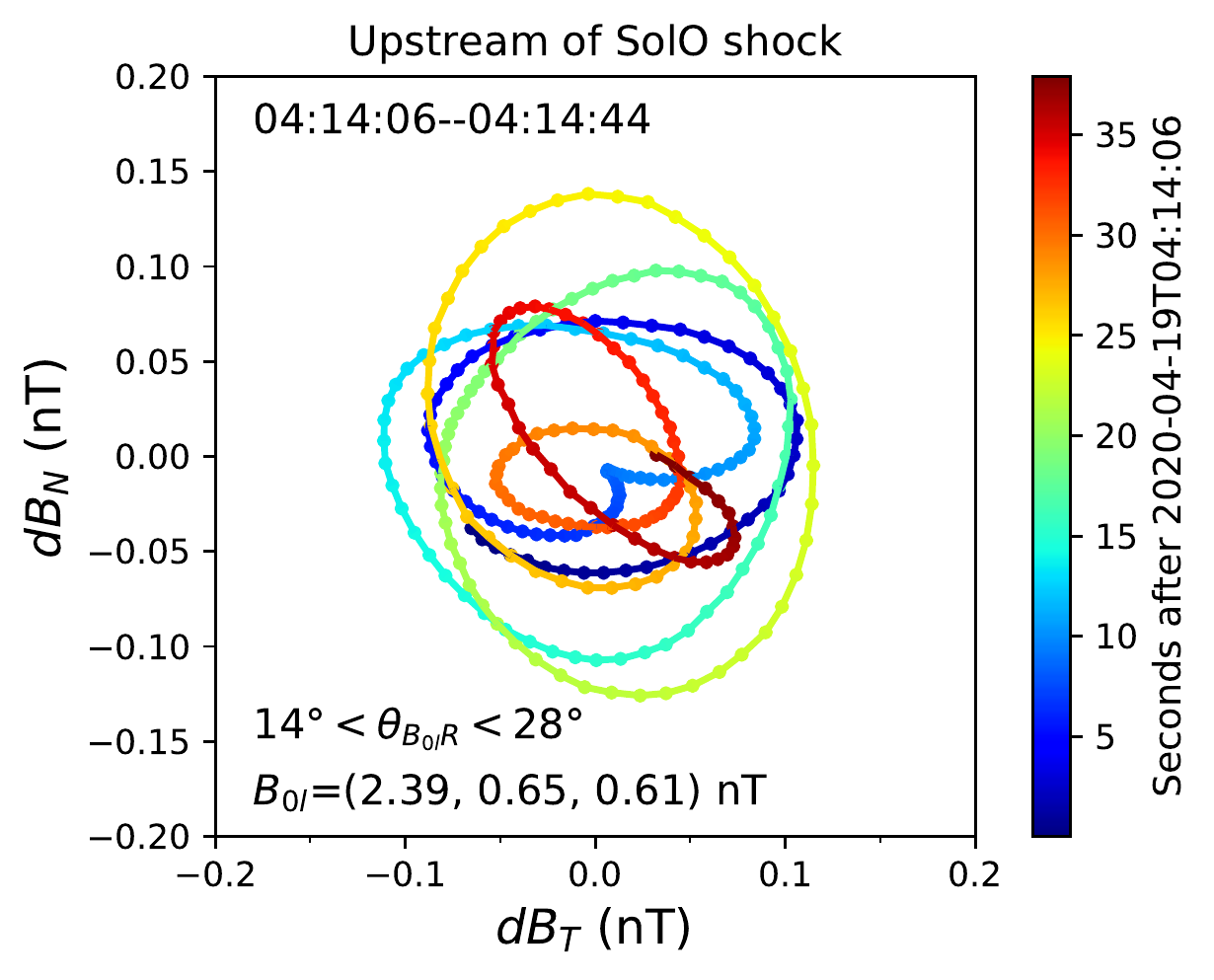}%
\includegraphics[width=0.5\linewidth]{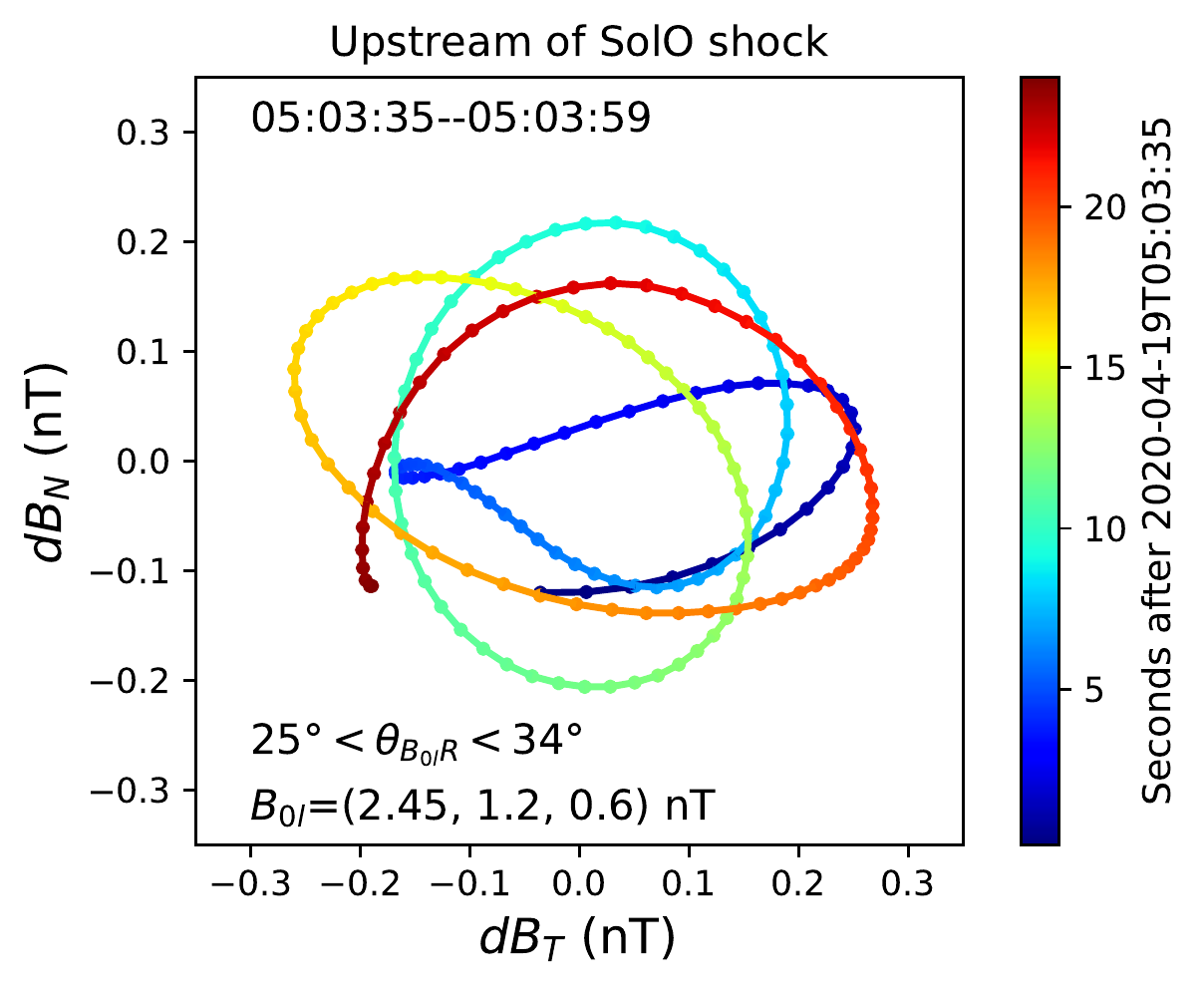} \\ 
	\caption{Two cases of the $dB_T$--$dB_N$ hodograms observed by \textit{Solar Orbiter} upstream of the shock. The magnetic field fluctuations $dB_T$ and $dB_N$ are reconstructed from the wavelet transform by averaging in the period range [3, 8] s. The first case is selected in the time interval between 04:14:06 UT and 04:14:44 UT on 2020 April 19 when $14\degr < \theta_\mathrm{kB} < 28\degr$. The second case is selected in the time interval between 05:03:35 UT and 05:03:59 UT on 2020 April 19 when $22\degr < \theta_\mathrm{kB} < 34\degr$. $B_\mathrm{0l}$ is the local mean magnetic field averaged in the selected period range and time interval.}\label{fig:solo-whistler}
\end{figure}

To further identify the wave modes, we analyze the hodograms of the $B_T$ and $B_N$ fluctuations obtained from wavelet decomposition. The method has been successfully applied to diagnose the kinetic waves in the solar wind \citep{he2011oblique}.
Figure \ref{fig:solo-whistler} shows two examples of the hodograph of the magnetic field fluctuations $dB_T$--$dB_N$ in \textit{Solar Orbiter}'s upstream region. The first interval extends from 04:14:06 UT to 04:14:44 UT on 2020 April 19. The second interval extends from 05:03:35 UT to 05:03:59 UT. Both intervals are chosen by the enhanced magnetic helicity shown in Figure \ref{fig:solo-shock}. The magnetic fluctuations $dB_T$ and $dB_N$ are reconstructed from the wavelet transform by averaging in the period range between 3 s and 8 s. Both intervals show clearly right-handed polarization ellipses in the T--N plane ($B_R > 0$). 
The local mean magnetic field $B_\mathrm{0l}$ for each interval is listed in the figure, which has a small tilt angle to the $R$ direction (assumed wavevector $k$ direction), i.e., the angle between $B_\mathrm{0l}$ and $R$ direction is around 20$\degr$ for the first interval, and about 29$\degr$ for the second interval. $B_\mathrm{0l}$ is almost perpendicular to the $T$--$N$ plane, indicating the dominance of perpendicular fluctuations $dB_{\perp}$. Due to the relatively low frequency (0.125--0.3 Hz) and relatively small magnetic compressibility ($C_B \simeq 0.04$), these signatures may indicate quasi-parallel Alfv\'en waves. However, we do not rule out the possibility of quasi-parallel fast-mode/whistler waves \citep{He2015}, which also possess the above properties. The exact wave mode identification needs further investigation. 

\begin{figure}[htbp]
\centering
\includegraphics[width=0.5\linewidth]{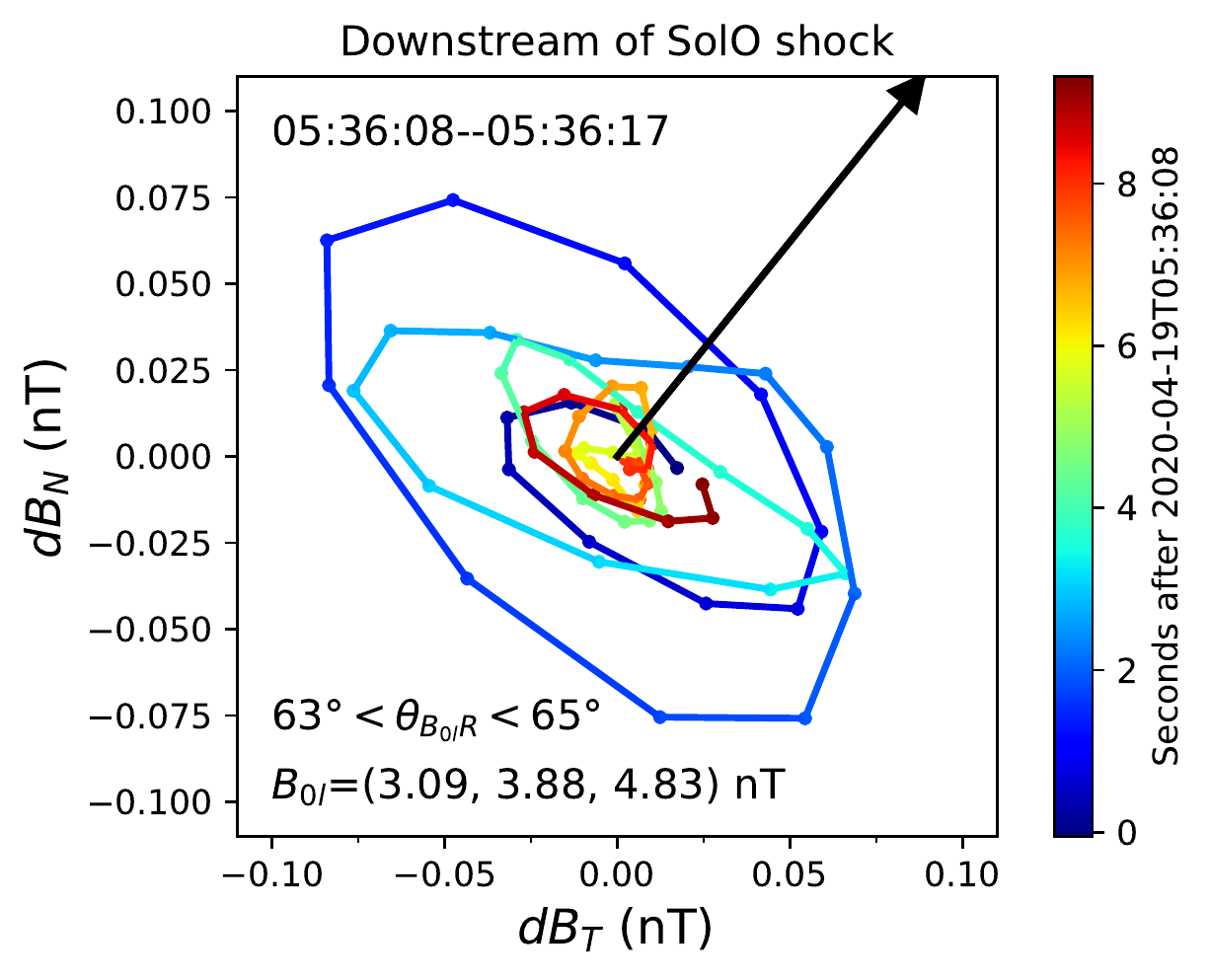}%
\includegraphics[width=0.5\linewidth]{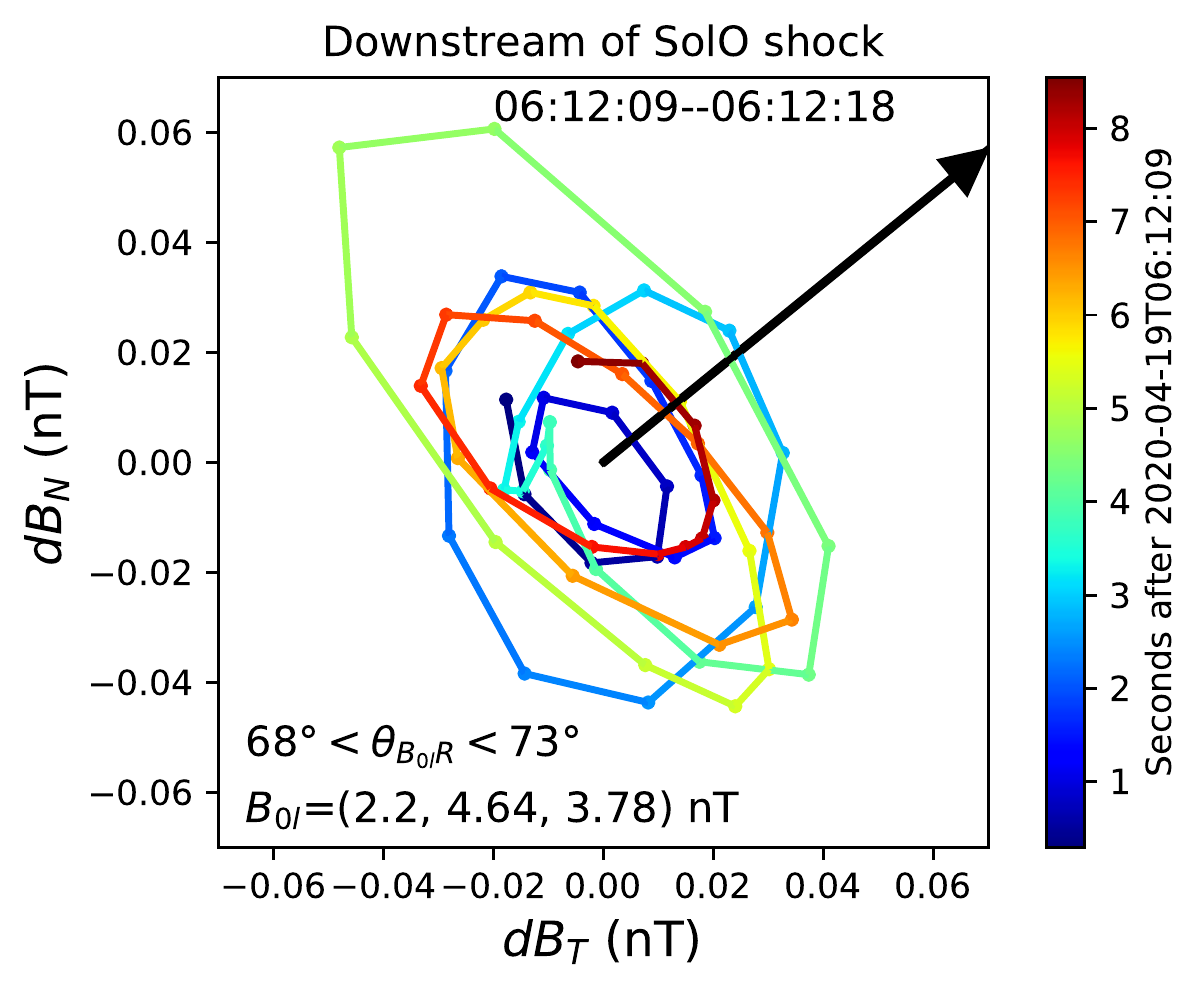} \\ 
	\caption{Magnetic hodograms in the same format as Fig.\ \ref{fig:solo-whistler}, but for \textit{Solar Orbiter} observations downstream of the shock. The wavelet reconstructed magnetic field fluctuations $dB_T$ and $dB_N$ are averaged over the period range [0.5, 2] s. The black arrow indicates the direction of the local mean magnetic field $B_\mathrm{0l}$. }\label{fig:solo-KAW}
\end{figure}

Similarly, the magnetic hodograms of the \textit{Solar Orbiter} downstream waves are presented in Figure \ref{fig:solo-KAW}. The magnetic field fluctuations $dB_T$ and $dB_N$ are obtained by averaging the wavelet decomposition in the period range [0.5, 2] s. Because of the increased wave frequency downstream, the averaging period is smaller than that used upstream. This is also reflected by the fewer data points on the ellipses compared to upstream. The local downstream mean magnetic field $B_\mathrm{0l}$ is quasi-perpendicular to the $R$ direction, being 63.5$\degr$ in the first interval and 70$\degr$ in the second interval. Their directions in the $T$--$N$ plane are shown as the black arrow in each panel.  
The two intervals show reasonably well defined polarization ellipses with the major axis perpendicular to the local mean magnetic field, indicating $dB_{\parallel}<dB_{\perp}$. The wave frequency is in the range of 0.5--2 Hz and the compressibility is about 0.24. These features are consistent with oblique KAWs since oblique whistler waves usually have $dB_{\parallel}>dB_{\perp}$ with the major axis of the polarization ellipse parallel to $B_\mathrm{0l}$ \citep{he2011oblique}.

\begin{figure}[htbp]
\centering
\includegraphics[width=0.5\linewidth]{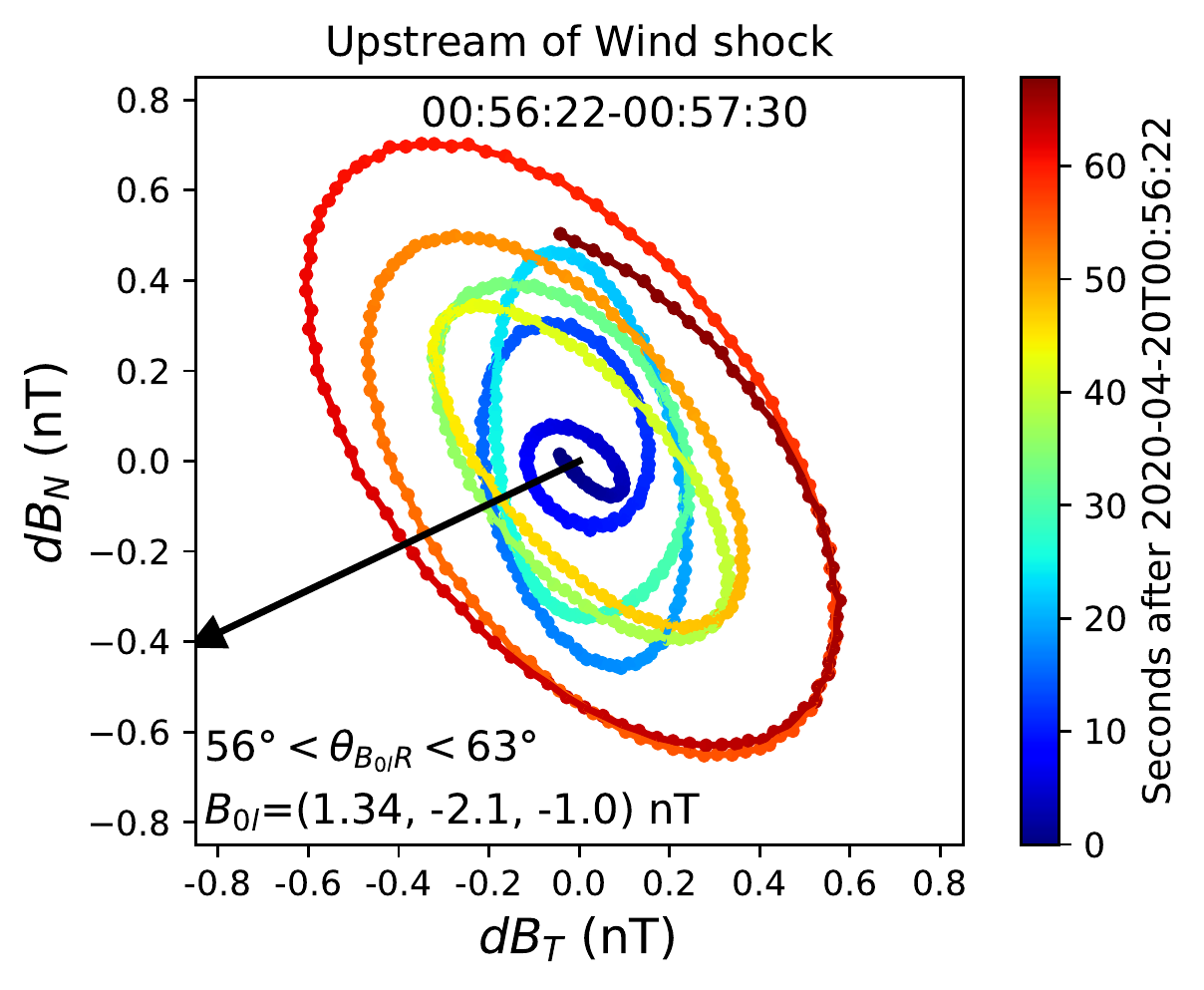}%
\includegraphics[width=0.5\linewidth]{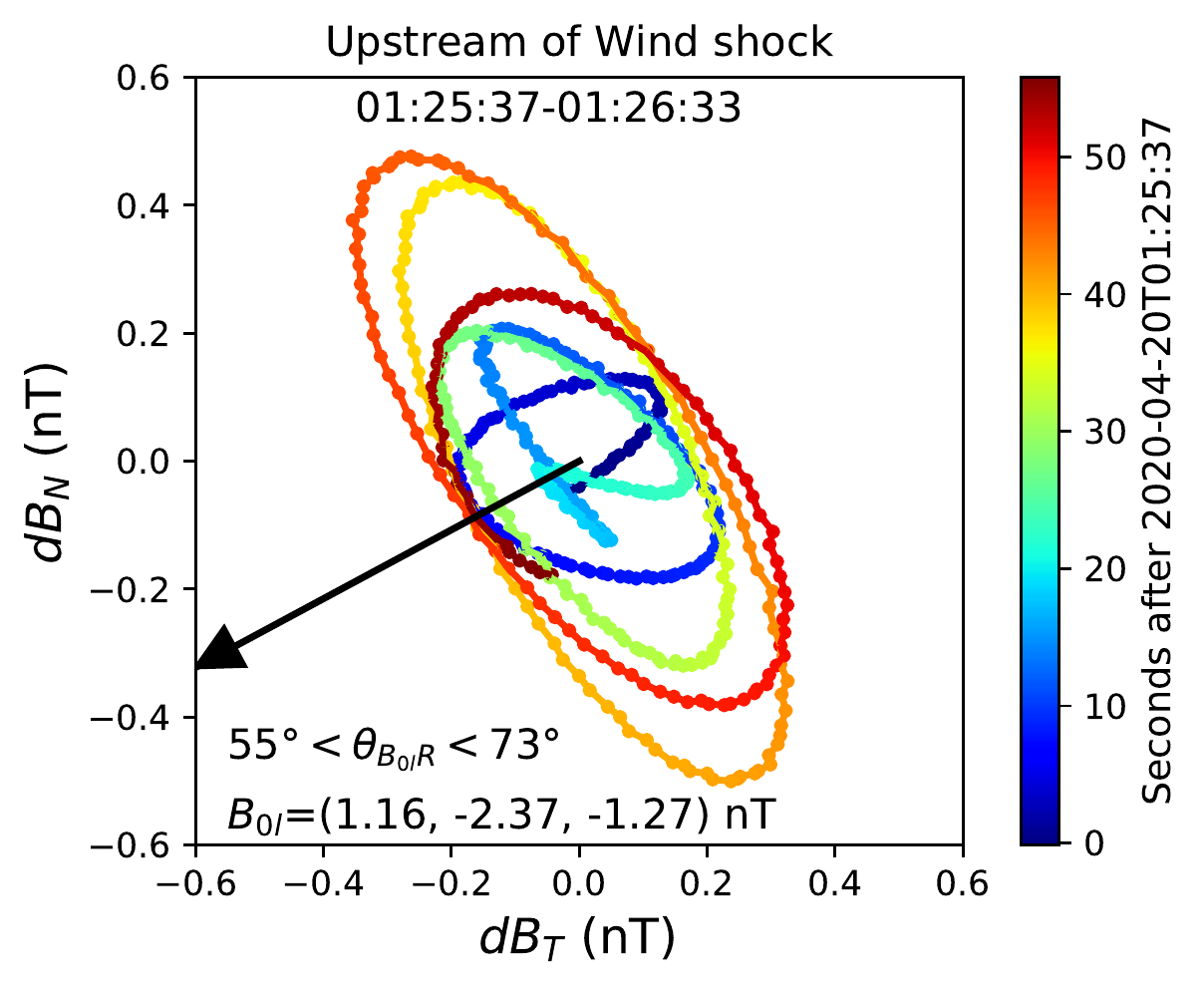} \\
\caption{Magnetic hodograms in the same format as Fig.\ \ref{fig:solo-whistler}, but for \textit{Wind} observations upstream of the shock. The wavelet reconstructed
	$dB_T$ and $dB_N$ are averaged over the period range [7, 10] s. Black arrow lines indicate the local mean magnetic field $B_\mathrm{0l}$ in the $T$--$N$ plane.}\label{fig:wind-KAW-up}
\end{figure}

\begin{figure}[htbp]
\centering
\includegraphics[width=0.5\linewidth]{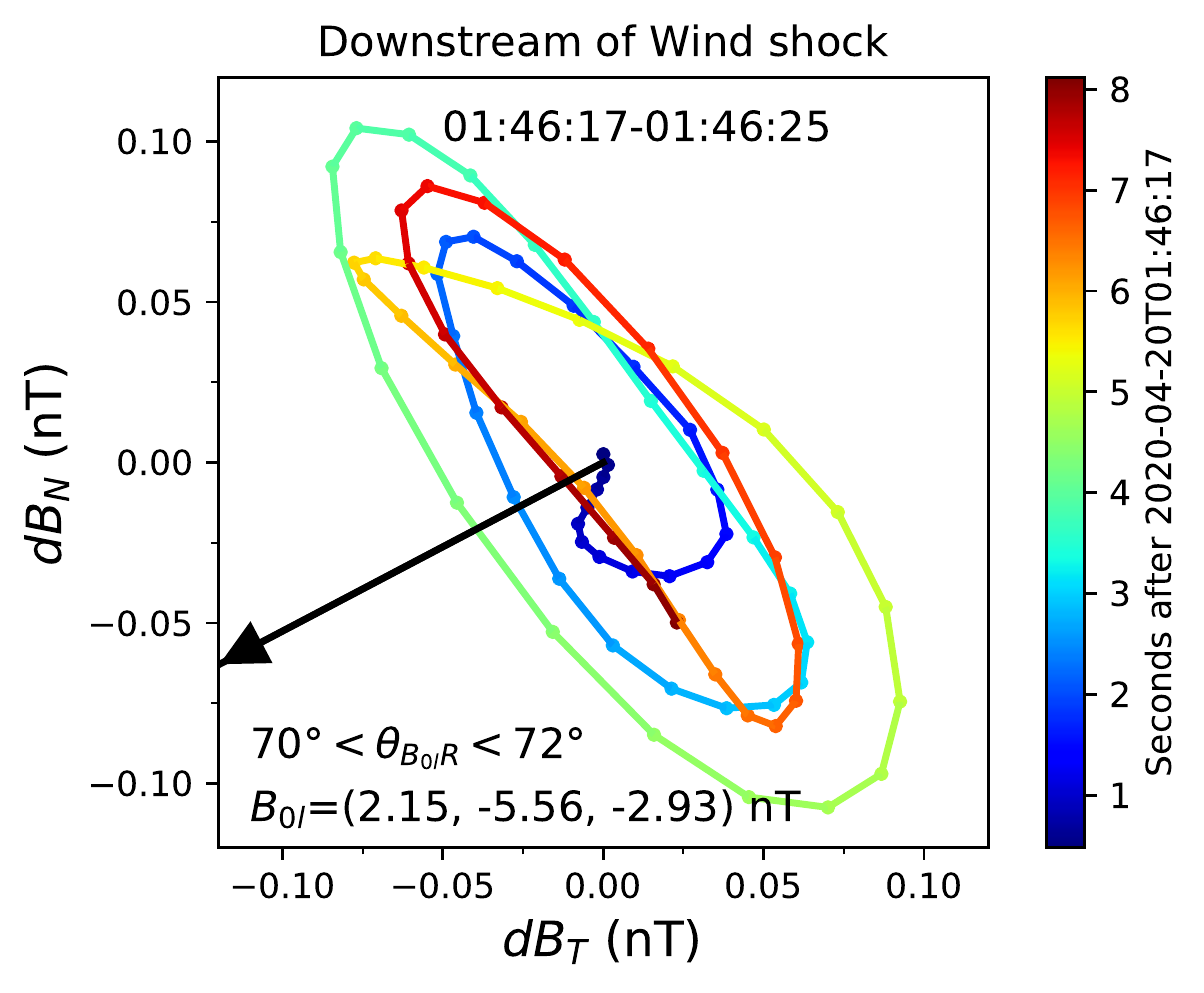}%
\includegraphics[width=0.5\linewidth]{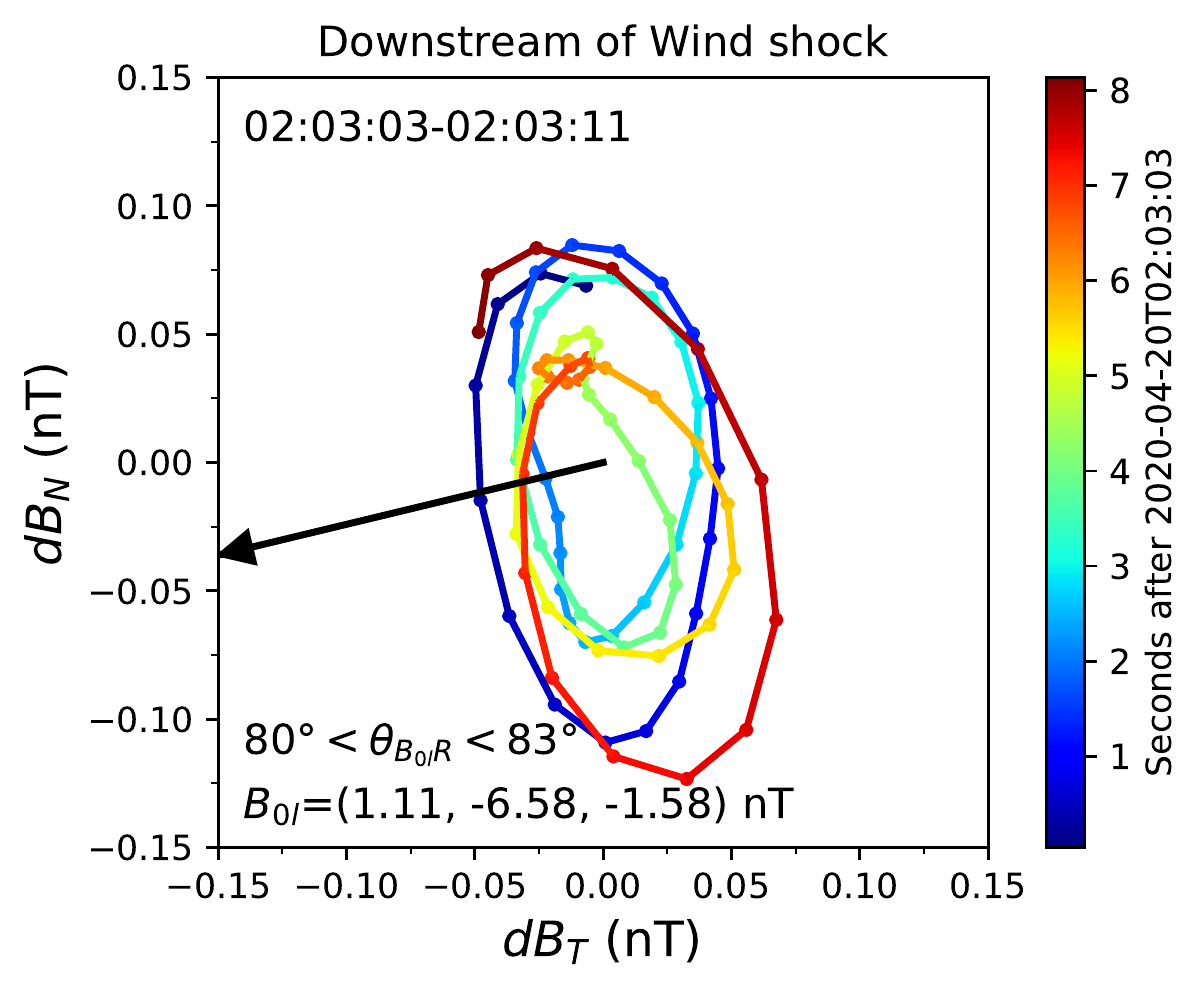}
\caption{Magnetic hodograms in the same format as Fig.\ \ref{fig:solo-whistler}, but for \textit{Wind} observations downstream of the shock. The reconstructed magnetic field fluctuations are averaged in the period range [0.7, 2] s. }\label{fig:wind-KAW}
\end{figure}

The same analysis is also performed for \textit{Wind} observations upstream and downstream of the shock. The results are shown in Figures \ref{fig:wind-KAW-up} and \ref{fig:wind-KAW}, respectively. The four intervals are again selected by the enhanced magnetic helicity $\sigma_m (s, t)$ shown in Figure \ref{fig:wind-shock}. All of these intervals show right-hand polarized ellipses in the $T$--$N$ plane ($B_R$>0) with the major axis perpendicular to the local mean magnetic field $B_\mathrm{0l}$. The upstream and downstream $\theta_\mathrm{kB}$ observed by \textit{Wind} is comparably larger than that observed by \textit{Solar Orbiter}, indicating that the wave modes here are more perpendicular with 50$\degr$<$\theta^u_\mathrm{kB}$<75$\degr$ upstream and 65$\degr$<$\theta^d_\mathrm{kB}$<90$\degr$ downstream. Similar to \textit{Solar Orbiter}, the upstream waves observed by \textit{Wind} also fall in a low frequency band (0.01--0.14 Hz) within the MHD inertial scale, a relatively small magnetic compressibility ($C_B\simeq0.05$), and a dominant perpendicular fluctuation $dB_{\perp}$, which can be oblique right-hand polarized Alfv\'en waves. Again, we do not exclude the possible existence of other MHD waves. The downstream wave frequency (0.5--1.4 Hz) is near the proton cyclotron frequency in the spacecraft frame and within the kinetic scale. The right-handed polarization with $dB_{\perp}>dB_{\parallel}$ suggests the oblique KAWs downstream, as also observed by \textit{Solar Orbiter}. 

\section{Summary and discussions}\label{sec:summary}

In conclusion, we have analyzed the properties of waves and turbulence near an ICME-driven shock observed by \textit{Solar Orbiter} at 0.8 AU and \textit{Wind} at 1 AU on 2020 April 19--20. The ICME is identified as a left-handed magnetic helical structure. The ICME-driven shock is a fast forward oblique shock with estimated speed $\sim$356 km/s at 1 AU. The shock obliquity at \textit{Wind}'s position is more perpendicular than that observed by \textit{Solar Orbiter}. The difference in the shock obliquity between \textit{Solar Orbiter} and \textit{Wind} may be due to the slight differences in their latitude and longitude and also to propagation effects. 
The main results are summarized as follows.
\begin{enumerate}
\item Spectral analysis of the magnetic field fluctuations show an enhanced fluctuating power in the shock downstream, suggesting that the shock can amplify the upstream turbulence as it is transmitted through the shock. This is consistent with theoretical expectations \citep{zank2021} and previous observations \citep{hu2013power, zhao2019, Borovsky2020JGR}. 
\item The total magnetic fluctuation power is dominated by the transverse fluctuations, which is consistent with nearly incompressible MHD turbulence models \citep{zank1992, zank1993, Zank2017, adhikari2017ii} and reported also in previous studies of downstream regions of interplanetary shocks \citep[e.g.,][]{moissard2019, good2020cross}. 
\item The magnetic compressibility $C_B$ is usually less than 0.1 in the inertial range but increases significantly as it approaches the kinetic range. The difference in upstream and downstream magnetic compressibility depends on the specific frequency range and the wave activity. For \textit{Wind} observations in the vicinity of the shock, the upstream $C_B$ is slightly larger than that downstream when the frequency exceeds 0.1 Hz. This also applies to \textit{Solar Orbiter}, but not for frequencies greater than 1 Hz.   
\item Both spacecraft observe upstream wave activity near the shock, which produced a clear bump in the magnetic field trace spectra and also the spectra of the normalized magnetic helicity $\sigma_m$. The bump is located near 0.1 Hz and is mostly due to the transverse fluctuations. Wave activity is also found in the downstream region, which can be transmitted from the upstream region and also can be locally generated. The frequency of the downstream wave increases by a factor of 7--10 due to the shock compression and Doppler effect.
\item The waves identified in this study are all right-hand polarized with positively enhanced $\sigma_m$. The hodograms of the magnetic fluctuations and $\sigma_m$ spectra indicate the existence of oblique kinetic Alfv\'en waves in the downstream region. The upstream waves observed by both spacecraft occur in a wide and low frequency range corresponding to ULF wave band, which can be low frequency Alfv\'en waves because of the small magnetic compressibility. However, we do not exclude the possibility of other low-frequency MHD waves being excited by ions and propagating upstream in the solar wind, such as the right-hand polarized fast mode wave, which needs further investigations. 
	
\end{enumerate}

Although we present evidence of wave activity using both spectral analysis and magnetic hodograms, the nature of the waves observed here is not conclusive. The relatively low frequency of the upstream waves suggests that they may not be associated directly with shock dissipation. Instead, they may be generated by the streaming of particles and contribute to the scattering and acceleration of energetic particles. The connection between these waves and particle acceleration remains to be understood and needs additional investigations in the future. 



\begin{acknowledgements}
	We acknowledge the partial support of the NSF EPSCoR RII-Track-1 Cooperative Agreement OIA-1655280 and a NASA award 80NSSC20K1783. D.T. is partially supported by the Italian Space Agency (ASI) under contract I/013/12/0. The Solar Orbiter magnetometer was funded by the UK Space Agency (grant ST/T001062/1). 	
\end{acknowledgements}

\bibpunct{(}{)}{;}{a}{}{,} 
\bibliographystyle{aa}
\bibliography{solorbiter}
\end{document}